\def\be{\begin{equation}}
\def\ee{\end{equation}}

\documentclass[aps,preprint,superscriptaddress,showpacs]{revtex4}
\usepackage{graphicx}
\usepackage{amssymb}
\usepackage{latexsym}

\begin{document}

\title{Dynamical heterogeneities and the breakdown\\ of the Stokes-Einstein
and Stokes-Einstein-Debye relations\\ in simulated water}
\date{\today}

\author{Marco G. Mazza}
\affiliation{Center for Polymer Studies and Department of Physics,
  Boston University, Boston, Massachusetts 02215, USA}
\author{Nicolas Giovambattista}
\affiliation{Department of Chemical Engineering, Princeton
 University,\\Princeton, New Jersey 08544-5263, USA}
\author{H. Eugene Stanley}
\affiliation{Center for Polymer Studies and Department of Physics,
  Boston University, Boston, Massachusetts 02215, USA}
\author{Francis  W. Starr}
\affiliation{Department of Physics, Wesleyan University, Middletown,
  Connecticut 06459, USA}

\begin{abstract}

We study the Stokes-Einstein (SE) and the Stokes-Einstein-Debye (SED)
relations, $D_t=k_BT/6\pi\eta R$ and $D_r=k_BT/8\pi\eta R^3$ where
$D_t$ and $D_r$ are translational and rotational diffusivity
respectively, $T$ is the temperature, $\eta$ the viscosity, $k_B$ the
Boltzmann constant and $R$ is the ``molecular'' radius, using
molecular dynamics simulations of the extended simple point charge
model of water. We find that both the SE and SED relations break down
at low temperature. To explore the relationship between these
breakdowns and dynamical heterogeneities (DH), we also calculate the
SE and SED relations for subsets of the $7\%$ ``fastest'' and $7\%$
``slowest'' molecules. We find that the SE and SED relations break
down in both subsets, and that the breakdowns occur on all scales of
mobility. Thus these breakdowns appear to be generalized phenomena, in
contrast with the view where only the most mobile molecules are the
origin of the breakdown of the SE and SED relations, embedded in an
inactive background where these relations hold. At low temperature,
the SE and SED relations in both subsets of molecules are replaced
with ``fractional'' SE and SED relations, $D_t\sim(\tau/T)^{-\xi_t}$
and $D_r\sim(\tau/T)^{-\xi_r}$ where $\xi_t\approx0.84<1$ and
$\xi_r\approx0.75<1$. We also find that there is a decoupling between
rotational and translational motion, and that this decoupling occurs
in both fastest and slowest subsets of molecules. We also find that
when the decoupling increases, upon cooling, the probability of a
molecule being classified as both translationally and rotationally
fastest also increases. To study the effect of time scale for SE and
SED breakdown and decoupling, we introduce a time-dependent version of
the SE and SED relations, and a time-dependent function that measures
the extent of decoupling. Our results suggest that both the decoupling
and SE and SED breakdowns are originated at the time scale
corresponding to the end of the cage regime, when diffusion
starts. This is also the time scale when the DH are more relevant.
Our work also demonstrates that selecting DH on the basis of
translational or rotational motion more strongly biases the
calculation of diffusion constants than other dynamical properties
such as relaxation times.
\end{abstract}
\maketitle

\section{\label{sec:intro}Introduction}

At temperatures where liquids have a diffusion constant similar to
that of ambient temperature water, the translational and rotational
diffusion, $D_t$ and $D_r$ respectively, are well described by the
Stokes-Einstein (SE) relation~\cite{einstein}
\be \label{eq:se}
D_t=\frac{k_BT}{6\pi\eta R}
\ee
and Stokes-Einstein-Debye (SED) relation~\cite{debye}
\be \label{eq:sed}
D_r=\frac{k_BT}{8\pi\eta R^3},
\ee
where $T$ is the temperature, $\eta$ the viscosity, $k_B$ the
Boltzmann constant and $R$ is the ``molecular'' radius.  These
equations are derived by a combination of classical hydrodynamics
(Stokes' Law) and simple kinetic theory (e.g, the Einstein
relation)~\cite{soft-matter}. Recently the limits of the SE and SED
relations have been an active field of
experimental~\cite{andreozzi,rossler,swallen,cicedi,chen-sed,chenbo},
theoretical~\cite{stillhod,lepo,ngai,kimkeyes,franosch,tarjus,coniglio,saltz}
and
computational~\cite{lepo2,netzbarb,matsui,heuer,bps,chand,kumdoug,zetter,bordat,kumarpnas,kbbsps,jung}
research. The general consensus is that the SE and SED relations hold
for low-molecular-weight liquids for $T\gtrsim 1.5 T_g$, where $T_g$
is the glass transition temperature. For $T\lesssim 1.5 T_g$,
deviations from either one or both of the SE and SED relations are
observed. Experimentally it is found that the SE relation holds for
many liquids in their stable and weakly supercooled regimes, but when
the liquid is deeply supercooled it overestimates $D_t$ relative to
$\eta$ by as much as two or three orders of magnitude, a phenomenon
usually referred to as the ``breakdown'' of the SE relation. The
situation for the SED relation is more complex. Some experimental
studies found agreement with the predicted values of the SED relation
even for deeply supercooled
liquids~\cite{cicedi,ediger-rev,sillescu-rev}, while others claim also
a breakdown of the SED relation to the same extent as for the SE
relation~\cite{chang,doss,steffen,fischer,rossler}.  The failure of
these relations provides a clear indication of a fundamental change in
the dynamics and relaxation of the system.  Indeed, the changing
dynamics of the liquid as it approaches the glass transition is well
documented, but not yet fully
understood~\cite{angell,nico-rev,ediger_glass,pablo}.

There is a large body of
evidence~\cite{weeks,bohmer,tracht,deschenes,deschenesJPC} that, upon
cooling, the liquid does not become a glass in a spatially homogeneous
fashion~\cite{richert,sillescu-rev,ediger-rev,glotzer-rev,kivFLD,ngai-PRE}.
Instead the system is characterized by the appearance of regions in
which the structural relaxation time can differ by orders of magnitude
from the average over the entire system~\cite{tamm}. The liquid is
characterized by the presence of ``dynamical heterogeneities'' (DH),
in which the motion of atoms or molecules is highly spatially
correlated.  The presence of these DH is argued to give rise to the
breakdown of the SE relation~\cite{stillhod,tarjus}.  Since the
derivation of the Einstein relation assumes uncorrelated motion of
particles, it is reasonable that the emergence of correlations could
result in a failure of the SE relation.  The aim of the present work
is to assess the validity of the SE and SED relations in the SPC/E
model of water, and consider to what extent the DH contribute to the
SE and SED breakdown.

Computer simulations have been particularly useful for studying DH
(e.g., see Refs.~\cite{hh,melcuk,pan,shell,kobdonati,nicoPRL}) since simulations
have direct access to the details of the molecular motion.  For
water, the existence of regions of enhanced or reduced mobility has
also been identified~\cite{nicoPRL,matharoo}. In particular,
Ref.~\cite{nicoPRL} identifies the clusters of molecules with greater
{\it translational} (or center of mass) mobility with the hypothesized
``cooperatively rearranging regions'' of the Adam-Gibbs
approach~\cite{AG,steve-wolynes}.

Most computer simulation studies on DH describe these heterogeneities
based on the particle or molecule translational degrees of freedom. We
will refer to these DH as translational heterogeneities (TH).  For
water, it is also necessary to consider the rotational degrees of
freedom of the molecule.  Recently, some computer simulation studies on
molecular systems described the DH based on the molecular {\it
  rotational} degrees of freedom
\cite{lepo2,KKS,andreozzi,matsui,kimkeyes1,kimlikeyes,first}.  We
will refer to these DH as rotational heterogeneities (RH). For the
case of a molecular model of water, RH were studied~\cite{first}
and it was found that RH and TH are spatially correlated.  This
work extends those results.  We find support for the idea that
TH are connected to the failure of the SE relation, and further that
RH have a similar effect on SED relation.  Additionally, we find
that the breakdown of these relations is accompanied by the
decoupling of the translational and rotational motion.

This work is organized as follows. In the next section we describe the
water model and simulation details.  In Sec.~\ref{sec:sesed} and
Sec.~\ref{sec:dh} we test the validity of the SE and SED relations and
their connection with the presence of DH, respectively. The decoupling
between rotational and translation motion is studied in
Sec.~\ref{sec:dec}. In Sec.~\ref{timeScaleSE-SED} we explore the role
of time scale in the breakdown of the SE and SED relations and
decoupling of rotational and translational motion. We summarize our
results in Sec.~\ref{sec:conc}. We have placed some technical aspects
of the work in appendices to facilitate the flow of our results.

\section{\label{sec:sim}Model and Simulation Method}

We perform molecular dynamics (MD) simulations of the SPC/E model of
water~\cite{spce}. This model assumes a rigid geometry for the water
molecule, with three interaction sites corresponding to the centers of
the hydrogen (H) and oxygen (O) atoms. Each hydrogen has a charge
$q_H=0.4238~e$, and the oxygen charge is $q_O=-2.0~q_H$, where $e$ is
the magnitude of the electron charge. The OH distance is $1.0$~\AA~and
the HOH angle is $109.47^{\circ}$, corresponding to the tetrahedral
angle. In addition to the Coulombic interactions, a Lennard-Jones
interaction is present between oxygen atoms of two different
molecules; the Lennard-Jones parameters are $\sigma=3.166$~\AA~and
$\epsilon=0.6502$~kJ/mol. We use a cutoff distance of
$2.5\sigma=7.915$~\AA~ for the pair interactions and the reaction
field technique~\cite{stein} is used to treat the long range Coulombic
interactions.

We perform simulations in the constant particle number, $N$, volume,
$V$, and temperature $NVT$ ensemble with $N=1728$ water molecules and
fixed density $\rho = 1.0$~g/cm$^3$.  The values of the simulated
temperature are $T=210$, $220$, $230$, $240$, $250$, $260$, $270$,
$280$, $290$, $300$, $310$, $330$~and~$350$~K. We use the Berendsen
method~\cite{berend} to keep the temperature constant. We use periodic
boundary conditions and a simulation time step of $1$~fs. To ensure
that simulations attain a steady-state equilibrium, we perform
equilibration simulations for at least the duration specified by
Ref.~\cite{francislong}. After these equilibration runs we continue
with production runs of equal duration during which we store the
coordinates of all atoms for data analysis.  To improve the statistics
of our results, we have performed $5$ independent simulations for each
$T$.  Ref.~\cite{francislong} provides further details of the
simulation protocol.

\section{\label{sec:sesed}Breakdown of the SE and SED relations}

To assess the validity of the SE and SED relations we
consider a simple rearrangement of Eqs.~(\ref{eq:se}) and
(\ref{eq:sed}), {\it i.e.} we define the SE ratio
\be
R_{\rm SE}\equiv\frac{D_t\eta}{T}
\label{eq:rse}
\ee
and the SED ratio
\be
R_{\rm SED}\equiv\frac{D_r\eta}{T}.
\label{eq:rsed}
\ee
Both $R_{\rm SE}$ and $R_{\rm SED}$ will be temperature-independent if
the SE and SED relations are valid.

To evaluate $R_{\rm SE}$ and $R_{\rm SED}$, we must first
calculate the appropriate diffusion constants.  Following normal
procedure, we define
\be
D_t\equiv\lim_{\Delta t\to\infty}\frac{1}{6\Delta t}
\langle r^2(\Delta t)\rangle.
\label{dt}
\ee
where $\langle r^2(\Delta t)\rangle$ is the translational mean
square displacement (MSD) of the oxygen atoms 
\be 
\langle r^2(\Delta t)\rangle \equiv \frac{1}{N}
\sum_{i=0}^N|\vec{r}_i(t')-\vec{r}_i(t)|^2. 
\label{eq:msd} 
\ee
Here, $\vec{r}_i(t)$ and $\vec{r}_i(t')$ are the positions of the
oxygen atom of molecule~\emph{i} at time $t$ and $t'$ respectively,
and $\Delta t=t'-t$. Analogously, we define the rotational diffusion
coefficient
\be
D_r\equiv\lim_{\Delta t\to\infty}\frac{1}{4\Delta t}
\langle\varphi^2(\Delta t)\rangle,
\label{dr}
\ee
where $\langle\varphi^2(\Delta t)\rangle$ is the rotational mean
square displacement (RMSD) for the vector rotational displacement
$\vec{\varphi}_i(\Delta t)$. Special care must be taken to calculate
$\langle\varphi^2(\Delta t)\rangle$ so that it is unbounded.  A
detailed discussion of this procedure is provided in
Appendix~\ref{sec:rotational-msd}.

We also need the viscosity $\eta$ to evaluate $R_{\rm SE}$ and $R_{\rm
SED}$. Unfortunately, accurate calculation of $\eta$ is
computationally challenging. A frequently employed approximation
exploits the fact that $\eta$ is proportional to the shear stress
relaxation time, $\tau_s$, via the infinite frequency shear modulus,
$G_\infty$, which is nearly $T$-independent~\cite{mcdonal}.
Additionally, we expect that $\tau_s$ (a ``collective property'')
should be nearly proportional to other collective relaxation times,
such as the relaxation time $\tau$ defined from the coherent
intermediate scattering function, $F(q,\Delta t)$, where $q$ is the
wave vector. Therefore, we substitute $\eta$ by $\tau$, which should
only affect the value and units of the constants in the $R_{\rm SE}$
and $R_{\rm SED}$.  For the purposes of our calculations, we define
$\tau$ by fitting $F(q,\Delta t)$ at long times with a ``stretched''
exponential
\be 
\label{eq:isf} 
F(q,\Delta t)\sim\exp{[-(\Delta t/\tau)^\beta]}, 
\ee 
where $0<\beta<1$, and we focus on the $q$ value corresponding to the
first peak in the static structure factor $S(q)$.

\begin{figure}
\centerline{
\includegraphics[scale=0.4]{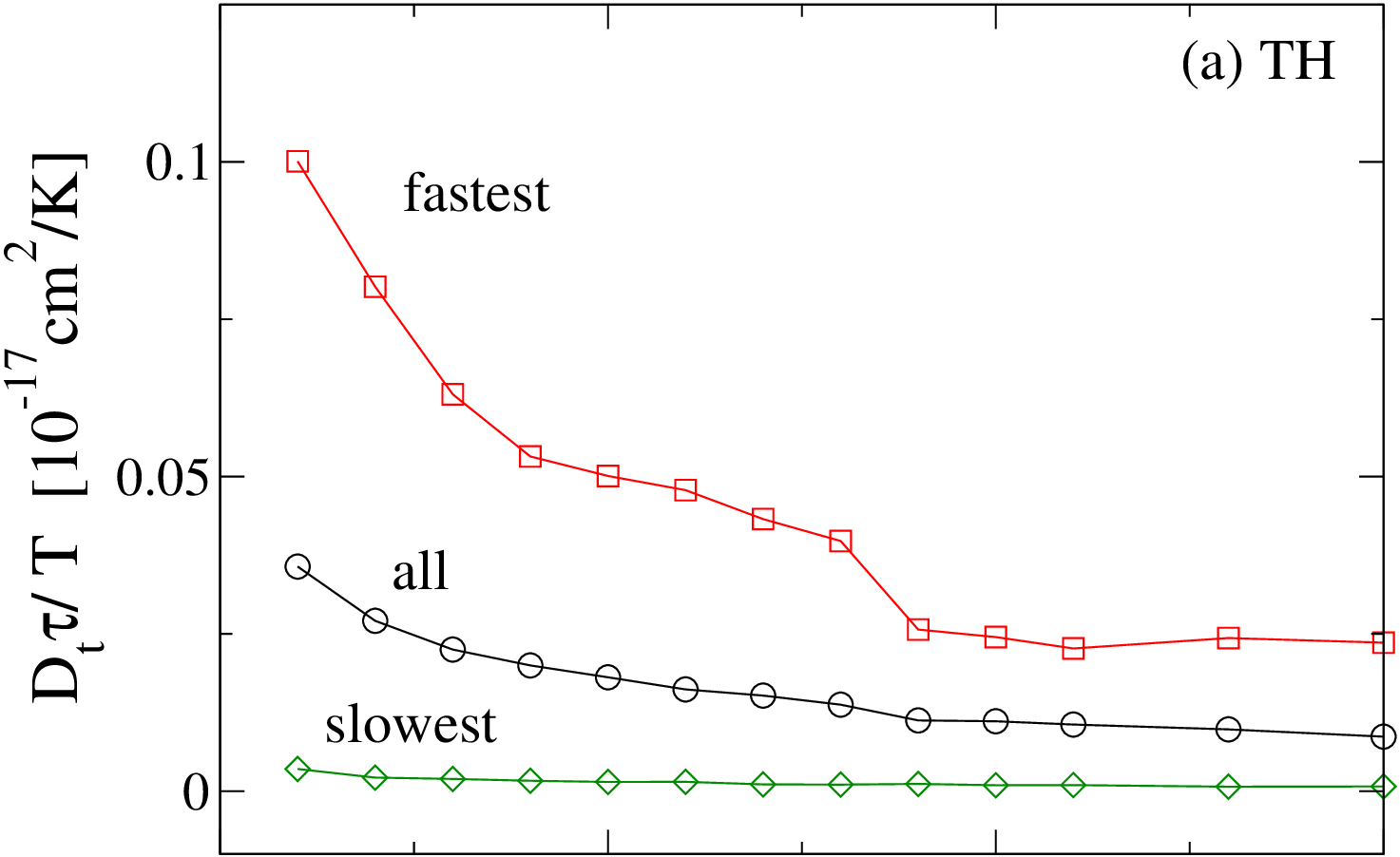}
\includegraphics[scale=0.4]{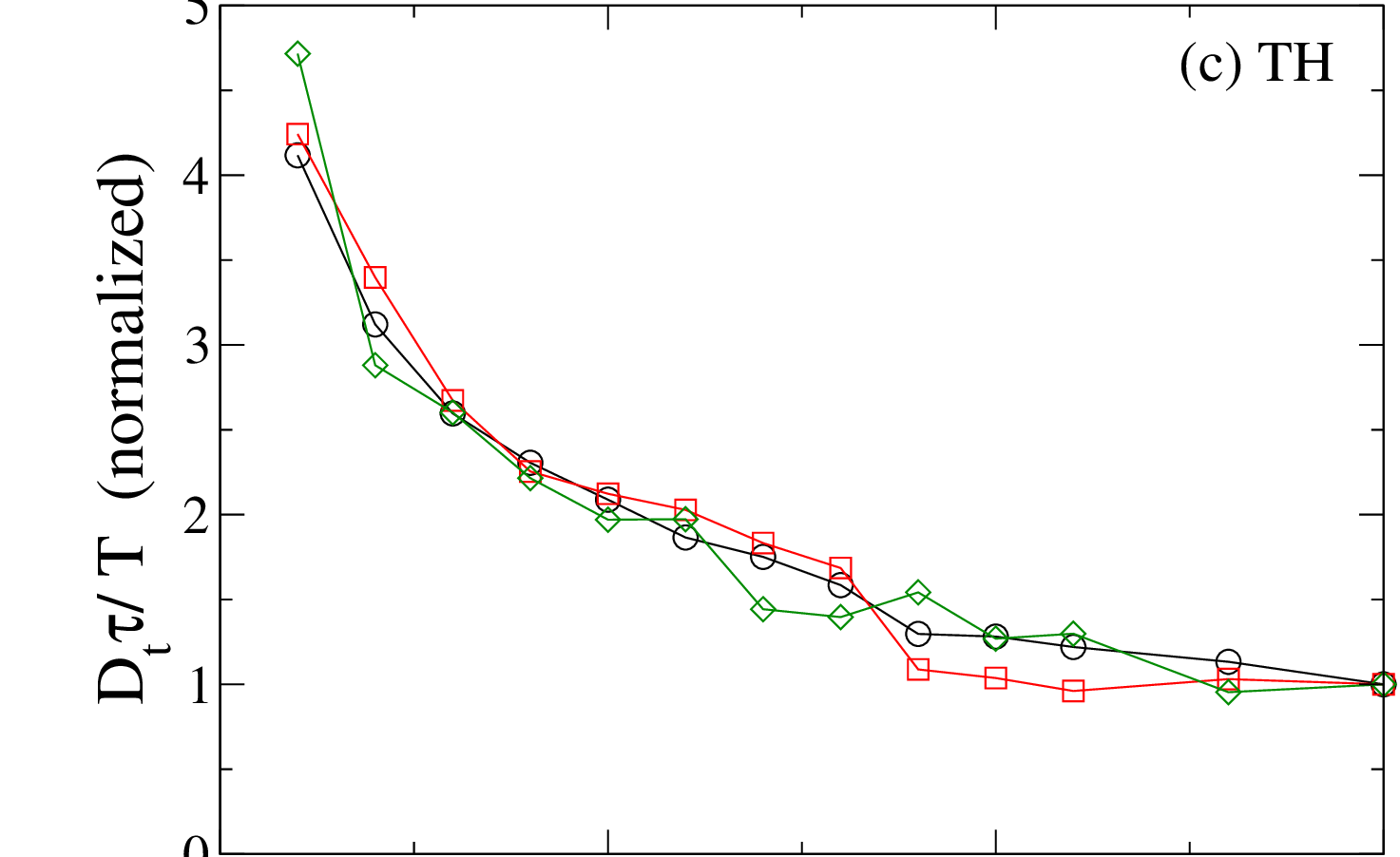}
}
\centerline{
\includegraphics[scale=0.4]{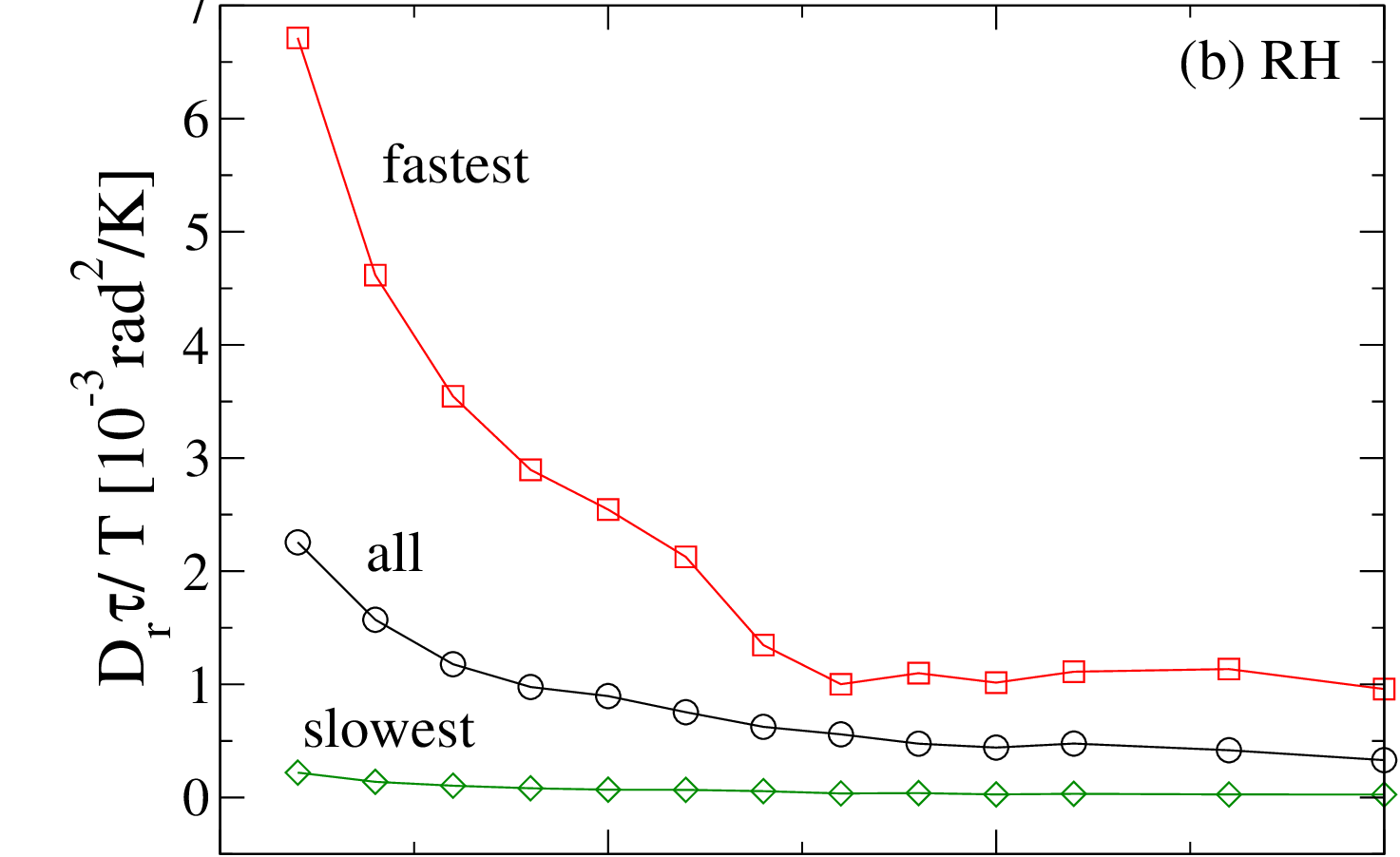}
\includegraphics[scale=0.4]{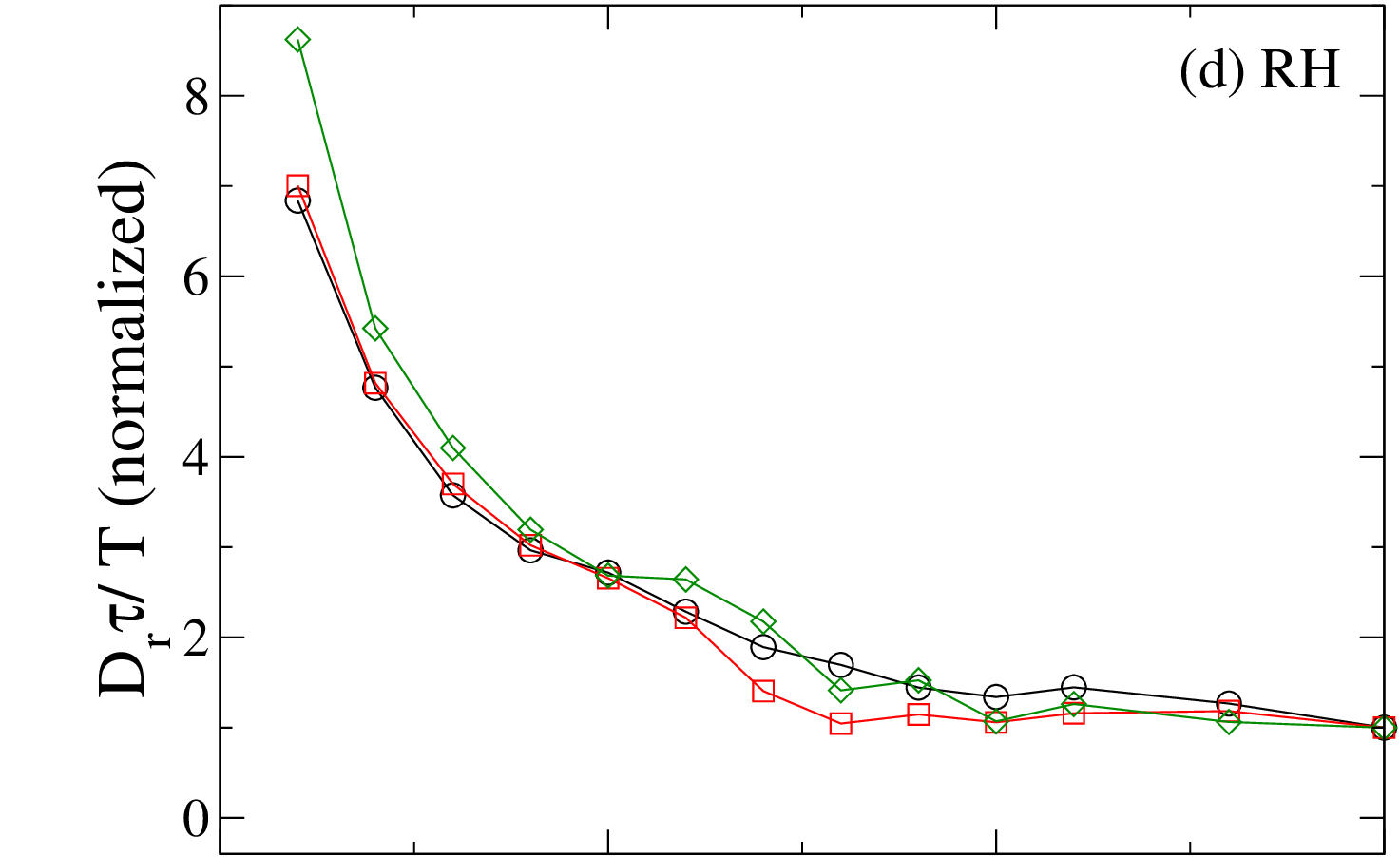}
} \caption{(a) SE ratio, $D_t\tau/T$, for the $7\%$ most
translationally mobile molecules (``fastest''), for the $7\%$ least
translationally mobile molecules (``slowest''), and for the entire
system (all). There is a breakdown of the SE relation (constant SE
ratio) at low temperatures in both the fastest and slowest subsets, as
well as for the entire system. (b) SED ratio, $D_r\tau/T$, for the
$7\%$ most rotationally mobile molecules, for the $7\%$ least
rotationally mobile molecules, and for the entire system
(all). Similar to (a), there is a breakdown of the SED relation
(constant SED ratio). (c) and (d) Normalization of the curves in (a)
and (b), respectively, by the corresponding quantities at
$T=350$~K. The collapse of these curves demonstrates that the relative
deviations from the SE and SED relations are approximately the same
for the corresponding mobility subsets. \label{DtauT} }
\end{figure}

Now that we have the necessary quantities, we show $R_{\rm SE}$ and
$R_{\rm SED}$ in Fig.~\ref{DtauT}(a) and Fig.~\ref{DtauT}(b) with the
curves labelled with ``all''.  Both quantities deviate at low $T$ from
the corresponding constant values reached at high temperature
indicating a breakdown of both the SE and SED relations.

Whether there is a breakdown of the SED relation in experiments is not
clear. While some experiments measuring dipole relaxation times show
that the SED relation holds down to the glass
transition~\cite{ediger-rev,sillescu-rev}, other
experiments~\cite{ludemann} show that the SED relation fails for low
$T$. Our simulations are in agreement with the breakdown of the SED
ratio observed in, e.g., Ref.~\cite{bps}.  Figures~\ref{DtauT}(a) and
\ref{DtauT}(b) also show $R_{\rm SE}$ and $R_{\rm SED}$ for different
subsets of molecules to examine the role played by DH. This is
discussed in the following Section.

\section{\label{sec:dh}Role of dynamical heterogeneities}
\subsection{Identifying mobility subsets}

Many theoretical approaches (e.g.~\cite{stillhod,tarjus}) attempt to
explain the breakdown of SE and/or SED in terms of DH.  To this end,
we must first describe the procedure used to select molecules whose
motion (or lack thereof) is spatially correlated.  A variety of
approaches have been used to probe the phenomenon of DH.  Here we use
one of the most common techniques: partitioning a system into mobility
groups based on their rotational or translational maximum
displacement.

For the TH, we define the translational mobility, $\mu_i$, of a
molecule $i$ at a given time $t_0$ and for an observation time $\Delta
t$, as the maximum displacement over the time interval $[t_0,t_0 +
\Delta t]$ of its oxygen atom
\be
\label{eq:mobi}
\mu_i(t_0,\Delta t)\equiv\max\{|\vec{r}_i(t+t_0)-\vec{r}_i(t_0)|
\,,
\,t_0\le t\le t_0+\Delta t\}.
\ee
For the RH, following~\cite{first}, we define a rotational mobility
that is analogous to the translational case. In analogy with
Eq.~(\ref{eq:mobi}), we define the rotational mobility at time $t_0$
with an observation time $\Delta t$ as
\be
\psi_i(t_0,\Delta t)\equiv
\max\{|\vec{\varphi}_i(t+t_0)-\vec{\varphi}_i(t_0)|\,, \,t_0\le
t\le t_0+\Delta t\}\,.
\ee
We identify the subsets of rotationally and translationally
``fastest'' molecules as the $7\%$ of the molecules with largest
$\psi_i$ and $\mu_i$, respectively. Analogously, we identify the
subsets of rotationally and translationally ``slowest'' molecules as
the $7\%$ of the molecules with smallest $\psi_i$ and $\mu_i$,
respectively.  The choice of $7\%$ is made to have a direct comparison
with the analysis of Ref.~\cite{nicoPRL,first}, but the qualitative
details of our work are unaffected by modest changes in this
percentage.  In the following, we will refer to these subsets of
molecules as TH and RH, fastest and slowest, depending on whether we
consider the top or the bottom of the distribution of mobilities. We
will see that comparing the fastest and the slowest molecules will
reveal new features of DH.

\subsection{SE and SED relations for fastest and slowest molecules}

Having identified subsets of highly mobile or immobile molecules, we
can calculate the ratios $R_{\rm SE}$ and $R_{\rm SED}$ by limiting
the evaluation of $D_t$, $D_r$ and $\tau$ to these subsets.  This is
relatively straightforward for the diffusion constants, since they
depend only on single molecule averages.  For $\tau$, the situation is
more complex since $F(q,\Delta t)$ includes cross-correlations between
molecules.  Hence we specialize the definition of $F(q,\Delta t)$ for
the TH and RH subsets by introducing a definition that captures the
cross-correlation within subsets and between a subset and rest of the
system.  We call this function $F_{\rm subset}(q,\Delta t)$, which we
discuss in detail in Appendix~\ref{sec:app2}.

We show the value of $R_{\rm SE}$ and $R_{\rm SED}$ in
Fig.~\ref{DtauT}(a) and~\ref{DtauT}(b) for the cases when only the
fastest and slowest subsets of molecules are considered.  Like the
total system average, both the SE and SED ratios for the subsets
deviate at low $T$ from the corresponding constant value reached at
high temperature. Therefore, we observe that the breakdowns of both
the SE and SED relations occur not only in the subset of the fastest
molecules, but also in the slowest. We have also confirmed a breakdown
in intermediate subsets.

The most mobile subset of molecules has a consistently greater value
of $D_t\tau /T$ and $D_r\tau /T$ than the rest of the system, while
the ratios for the least mobile subsets are always smaller. This is a
result of the fact that the means by which we select the different
subsets most strongly affects the diffusion constant (see
Appendix~\ref{sec:app2}), and hence the differences in the SE and SED
ratios between the full system and the subsets are dominated by the
diffusion constant, rather than by the relaxation time.

In order to compare the relative deviations of these curves from the
SE and SED predictions, we normalize $R_{\rm SE}$ and $R_{\rm SED}$ by
their respective high temperature values [Fig.~\ref{DtauT}(c)
and~\ref{DtauT}(d)]. We observe that there is a collapse of all the
curves; thus, we conclude that both the most and least mobile
molecules contribute in the same fashion to the breakdown of SE and
SED. Moreover, this result supports the scenario that the deviation
from the SE and SED relations cannot be attributed to only one
particular subset of fastest/slowest molecules, but to all scales of
translational and rotational mobility.  We have confirmed this by
looking at subsets of intermediate mobility (not shown).  Therefore,
we interpret our results as a sign of a generalized breakdown in the
system under study, in contrast to a picture where only the most
mobile molecules are the origin of the breakdown of SE and SED,
embedded in an inactive background where the SE and SED equations
hold~(see e.g.~\cite{cicedi}).  These results are consistent with the
results of Ref.~\cite{bps}, who arrived at the same conclusion via a
different analysis.

\subsection{Fractional SE and SED relations}

\begin{figure}
\begin{center}
\includegraphics[scale=0.45]{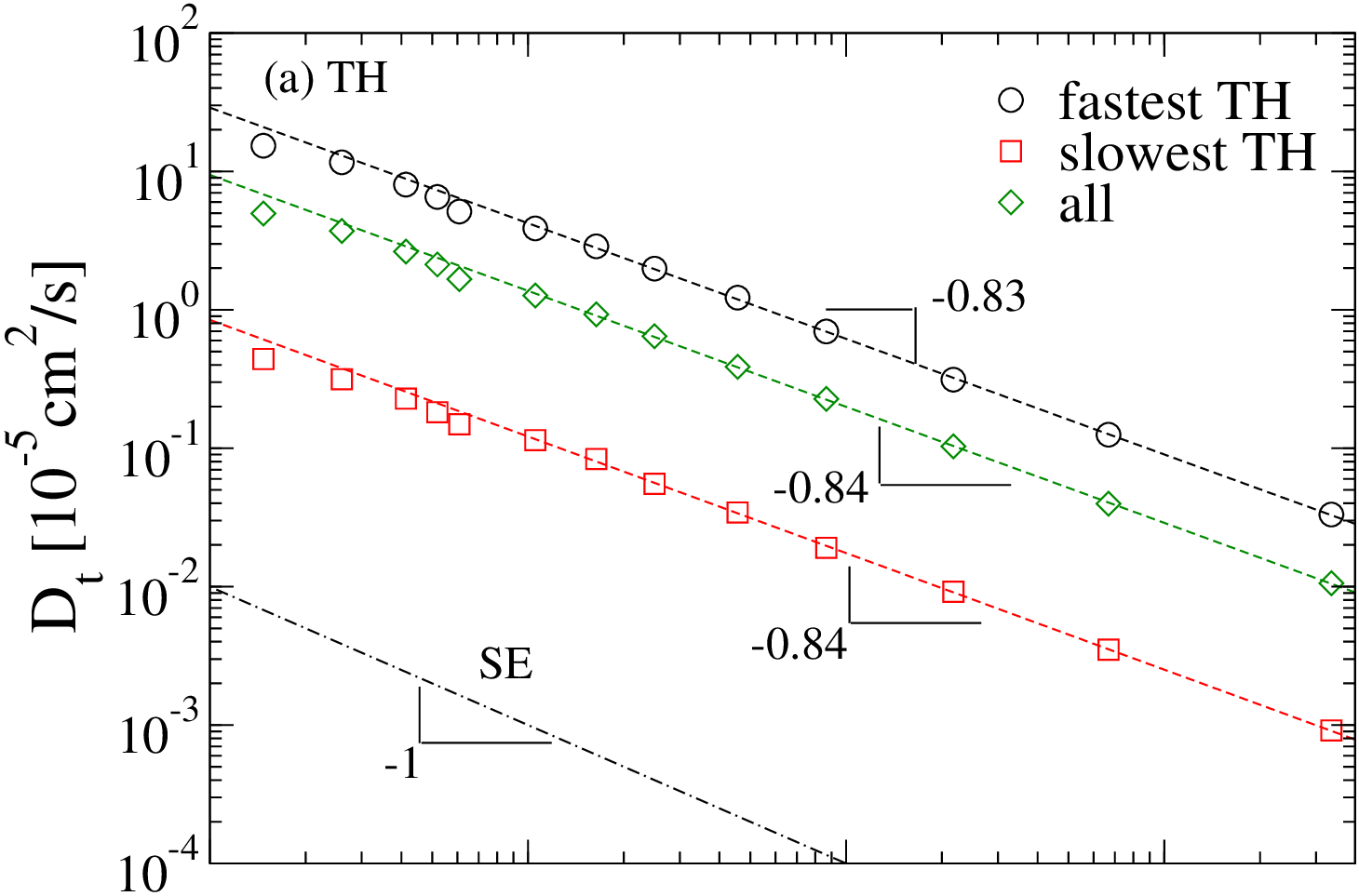}
\includegraphics[scale=0.45]{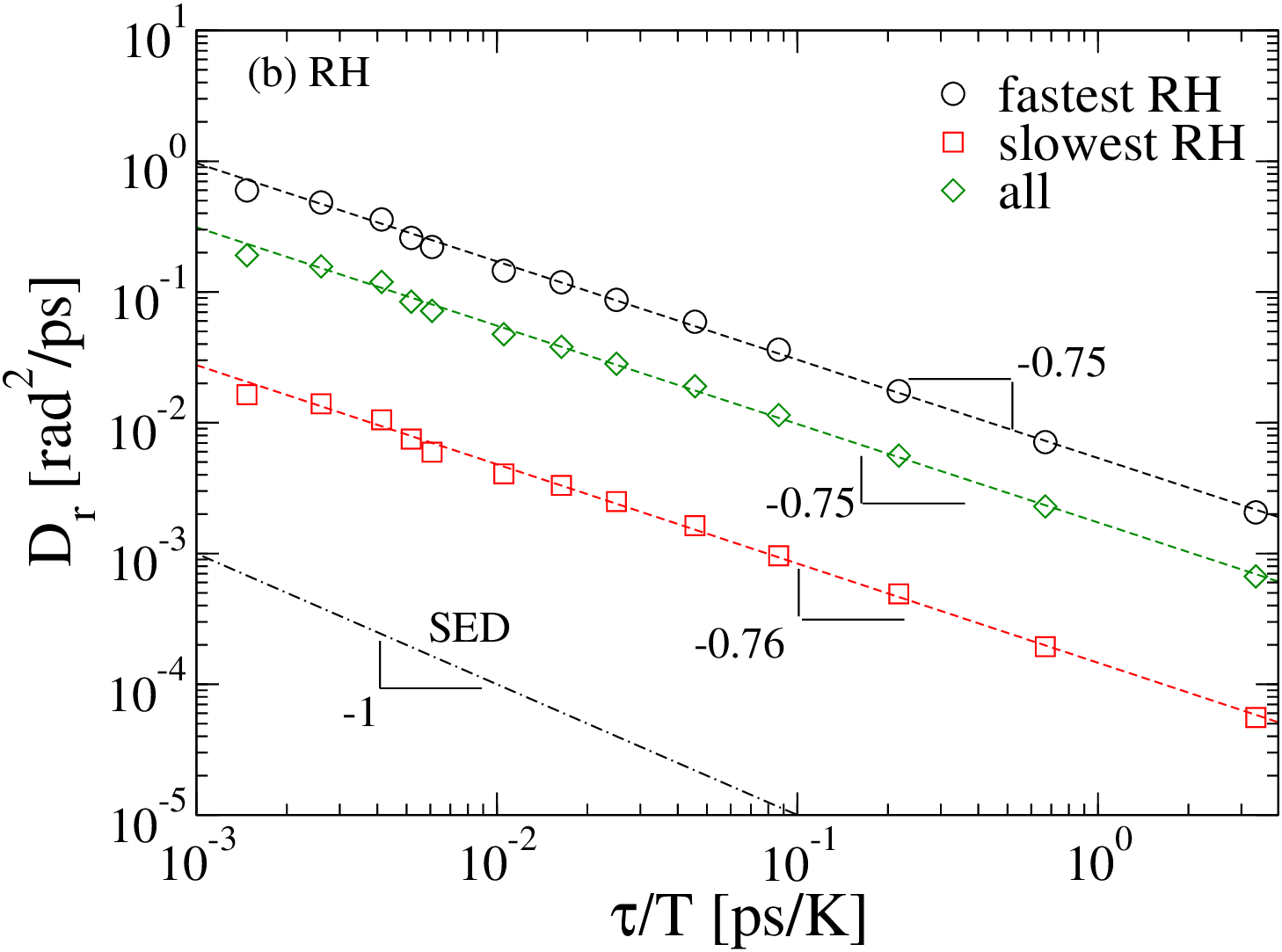}
\end{center}
\caption{(a) Power law fits of translational diffusivity $D_t$ as functions
of $\tau/T$, $D_t\sim(\tau/T)^{-\xi_t}$, for the eight values of
temperature $T=210$ $\dots$ $280$~K (but not for the remaining values
$T=290$ $\dots$ $350$~K), for fastest TH, slowest TH, and all
molecules. We estimate $\xi_t\approx0.84$. The dot-dashed line
represents the normal SE behavior ($\xi_t=1$). Consistently with the
results of Fig.~\ref{DtauT}, the deviation of these three curves from
the SE behavior is almost identical as reflected in the values of
these fractional exponents. (b) Power law fits of rotational
diffusivity, $D_r$, as functions of $\tau/T$,
$D_r\sim(\tau/T)^{-\xi_r}$, of simulations in the same temperature
range of (a) for fastest RH, slowest RH, and all molecules. We
estimate $\xi_r\approx0.75$. The dot-dashed line represents the normal
SED behavior ($\xi_r=1$). Also for RH, a fractional law is found with
the same exponents for the three families considered, and, noticeably,
the deviation from the normal case ($\xi_r=1$), is stronger for $D_r$
than for $D_t$, since $\xi_r<\xi_t$. \label{fractional}}
\end{figure}

When the SE and SED relations fail, it is frequently observed that
they can be replaced by \emph{fractional} functional
forms~\cite{poll,cicedi96,swallen,boc,vor,chang,andreozzi2,andreozzi,biel}
\begin{equation}
\label{eq:frac}
D_t\sim\left({\tau\over T}\right)^{-\xi_t},\quad 
D_r\sim\left({\tau\over T}\right)^{-\xi_r}
\end{equation}
with $\xi_t<1$ and $\xi_r<1$. Hence we test to what degree
Eqs.~(\ref{eq:frac}) hold for our system.  In Fig.~\ref{fractional} we
show a parametric plot of diffusivity versus $\tau/T$ for the entire
system, and for the fastest and slowest molecules composing the TH and
RH.  The results at low temperature are well fit with the fractional
form of SE and SED relations. From Fig.~\ref{fractional}, $\xi_t$ for
TH is $0.83, 0.84, 0.84$ for fastest, slowest, and all, respectively,
so all TH have approximately the same exponent. Similarly, for RH we
find that $\xi_r$ is $0.75, 0.76, 0.75$ for fastest, slowest, and all,
respectively.

Reference~\cite{bps} found a stronger form of this fractional
relation.  Specifically, Ref.~\cite{bps} examined an ``ensemble'' of
systems of the ST2 water model at the same $T$, which by statistical
variation have fluctuations in the SE and SED ratios.  Nonetheless,
all systems collapsed to the same master curve when plotted in the
parametric form shown in Fig.~\ref{fractional}, meaning that the
systems dominated by mobile or immobile molecules collapse to the same
curve. While Ref.~\cite{bps} employed a very different method (small
systems followed for shorter times), the conclusion of our
Fig.~\ref{fractional} is the same: a generalized deviation from SE and
SED. However, Fig.~\ref{fractional} clearly shows that we do not find
a general collapse in our present calculation. To understand why, we
return to the fact that the method by which we define mobility affects
much more strongly the diffusion constants than the coherent
relaxation time, $\tau$.  As a result, it is impossible to have the
data for the mobile and immobile subsets to collapse to a single
master curve. To observe the same collapse, presumably one needs a
more ``neutral'' method for selecting the mobile particles---that is
one that does not explicitly bias toward a specific
property. Unfortunately, such an approach is not obvious.  However, we
reproduced the ensemble approach of Ref.~\cite{bps}, by splitting each
of our 5 simulations into 3 trajectories. We obtain reasonable
fluctuations that allow us to test and confirm (not shown) the
observation of collapse of Ref.~\cite{bps}.  Hence, the phenomenon of
homogeneous breakdown of SE and SED appears to be robust for the
different water models.

\section{\label{sec:dec}Decoupling of Translational and Rotational
Motions}

The SE and SED relations also imply a coupling between rotational and
translational motion.  Specifically, Eqs.~(\ref{eq:se}) and
(\ref{eq:sed}) imply that the ratio
\be
\frac{D_r}{D_t}=\frac{3}{4R^2}
\ee
should remain constant as a function of temperature. Since we have
already seen that the SE and SED ratios are not obeyed, it is likely
that the ratio $D_r/D_t$ is also violated~\cite{ber}. However, it is also
possible that \textbf{$D_r/D_t$} remains constant if both $D_r$ and
$D_t$ deviate from their expected behavior in the same way.

\begin{figure}
\begin{center}
\includegraphics[scale=0.5]{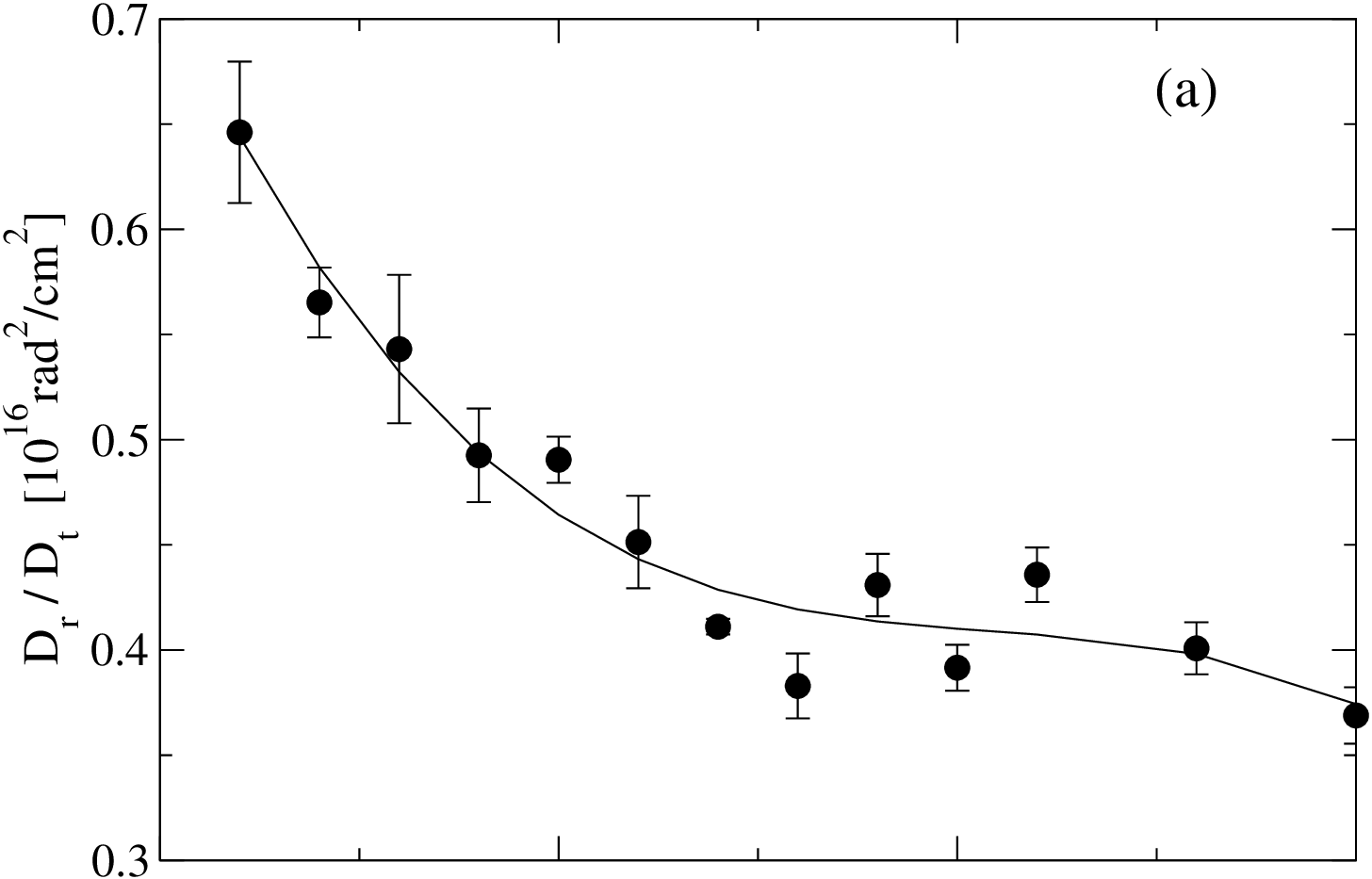}
\includegraphics[scale=0.5]{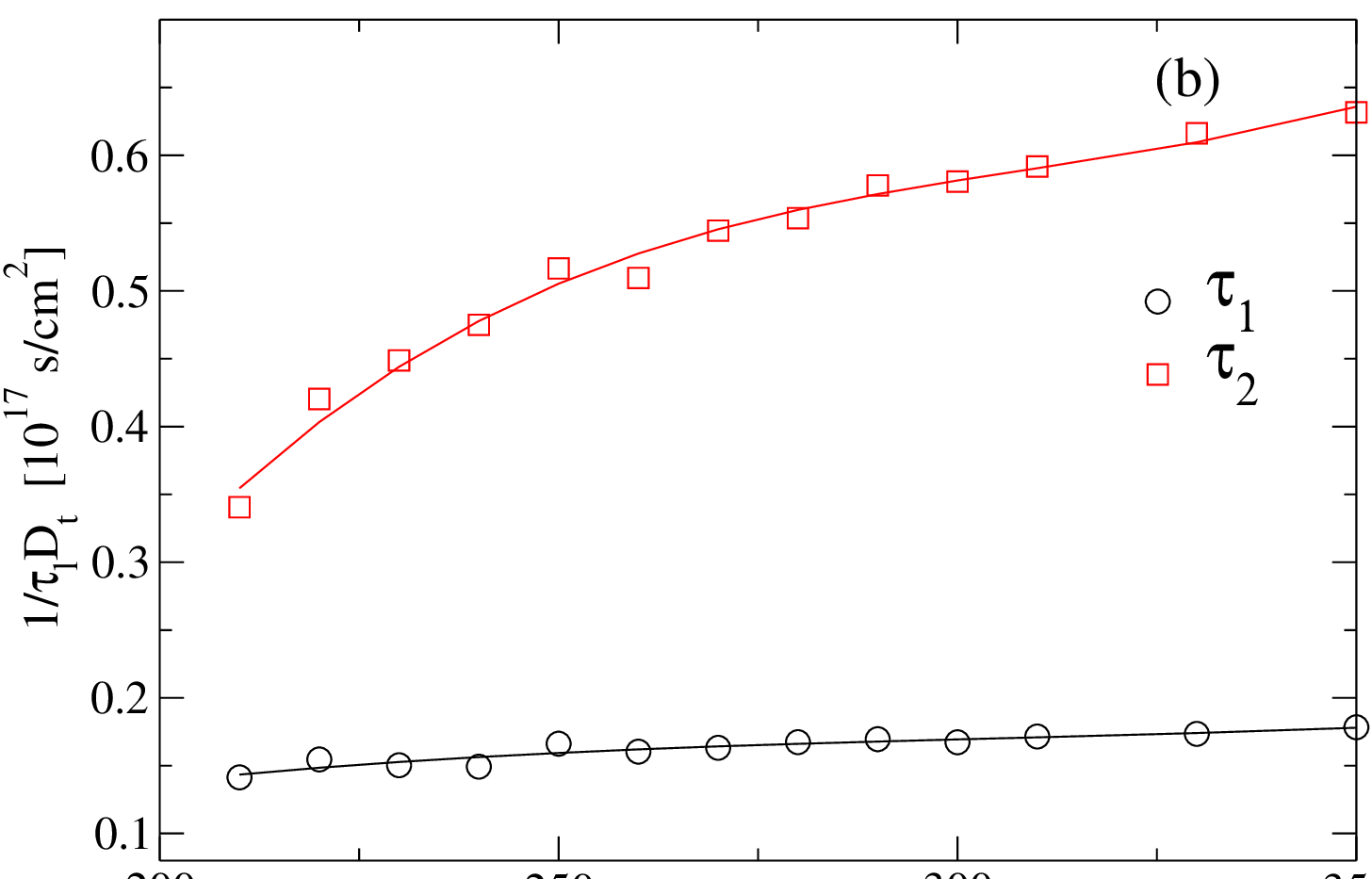}
\end{center}
\caption{(a) The ratio of rotational and translational diffusivities
$D_r/D_t$ as a function of temperature. As temperature decreases, this
ratio increases indicating a decoupling between rotation and
translational motion. The deviation of $D_r$ is stronger than that of
$D_t$. The line is a guide for the eye. (b) Same as (a) where the
rotational diffusivity, $D_r$, is replaced by the inverse of the
rotational relaxation time, $\tau_\ell$ with $\ell=1,2$, as usually
done in experiments. An opposite decoupling is observed in (a) and
(b). The lines are guides for the eye.\label{ratio}}
\end{figure}

Figure~\ref{ratio}(a) shows $D_r/D_t$ as a function of temperature.  As
$T$ decreases, we observe that $D_r/D_t$ increases, which implies that
the breakdown of the SED relation is more pronounced than that of the SE
relation.

Experiments generally do not examine the behavior of $D_r/D_t$ since
$D_r$ is not accessible.  Instead, $D_r$ is usually replaced by
$(\tau_\ell)^{-1}$ with $\ell=2$~\cite{netzbarb}. Here, $\tau_\ell$ is
the relaxation time of the rotational correlation function
\be
C_\ell(\Delta t) =\langle 
P_\ell(\cos[\hat{p}(t)\cdot\hat{p}(t+\Delta t)])\rangle,
\ee
where $P_\ell(x)$ is the Legendre polynomial of order $\ell$, and
$\hat{p}(t)$ is defined in Appendix~\ref{sec:rotational-msd}.
Figure~\ref{ratio}(b) shows $(\tau_\ell)^{-1}/D_t$ for $\ell=1,2$.  We
observe that $(\tau_\ell)^{-1}/D_t$ also shows a decoupling between
rotational and translational motion. However, while $D_r/D_t$
\emph{increases} upon cooling, $(\tau_\ell)^{-1}/D_t$
\emph{decreases} upon cooling. MD simulations using an
\emph{ortho}-terphenyl (OTP) model~\cite{OTP} and the ST2 water
model~\cite{bps} also find an opposite decoupling of the SE and SED
relations depending on whether $D_r$ or $\tau_2$ is used.  In the
simulations of OTP, it was shown that the inverse relation between
$D_r$ and $\tau_2$ fails due to the caging of the rotational motion;
this caging results in intermittent large rotations that are not
accounted for by the Debye approximation.

\begin{figure}
\begin{center}
\includegraphics[scale=0.5]{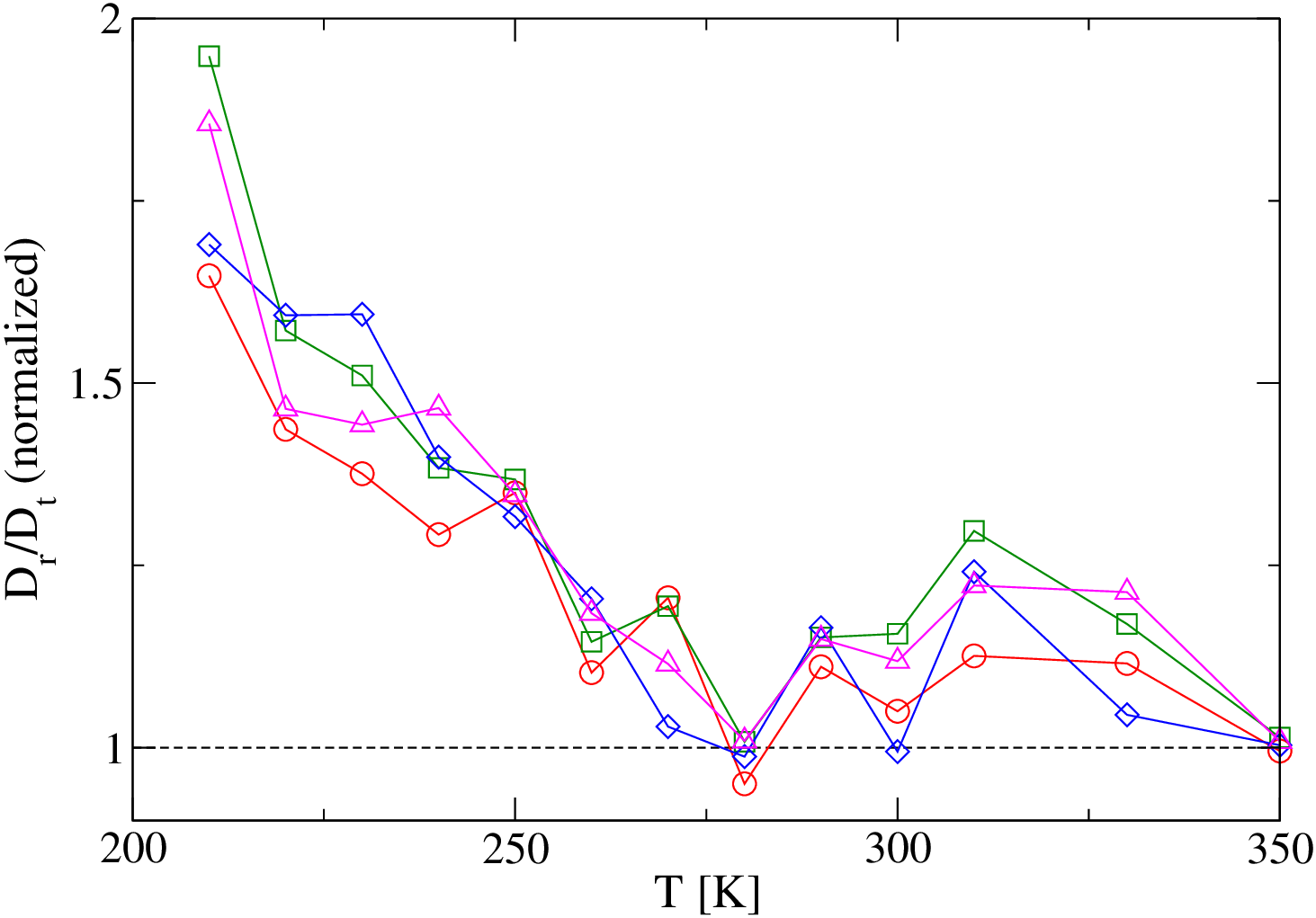}
\end{center}
\caption{The ratio of rotational and translational diffusivities,
$D_r$ and $D_t$ respectively, for the following choices of subsets:
$D_r$ for fastest TH divided by $D_t$ for fastest TH ($\lozenge$),
$D_r$ for slowest TH divided by $D_t$ for slowest TH ($\vartriangle$),
$D_r$ for fastest RH divided by $D_t$ for fastest RH ($\bigcirc$),
$D_r$ for slowest RH divided by $D_t$ for slowest RH ($\square$). The
values were normalized by the $T=350$~K values for every curve. The
fact that for these four cases $D_r/D_t$ deviates from unity (dashed
line) to approximately the same degree indicates that the decoupling
occurs across all four cases.\label{ratioDH}}
\end{figure}

Similar to the analysis of the breakdown of the SE and SED ratios, we
can test whether DH play a strong role in the decoupling by examining
the ratio $D_r/D_t$ for the different mobility subsets. This is
slightly complicated by the fact that we can choose mixed mobility
subsets when calculating the ratio.  Figure~\ref{ratioDH} shows that
the ratio $D_r/D_t$ for all choices of mobility subsets approximately
coincide when scaled by the high temperature behavior of $D_r/D_t$.
This indicates that (like the breakdown of the SE and SED relations)
the decoupling is uniform across the subsets of mobility.

\section{\label{timeScaleSE-SED}Time scales for breakdown and decoupling}
\subsection{Time dependent SE and SED relations}

The SE and SED relations depend on $D$ and $\eta$, which are defined
only in the asymptotic limit of infinite time. In contrast, the time
scale on which DH exist is finite, and generally shorter that the time
scale on which the system becomes diffusive.  As a result, making the
connection between DH and the breakdown of SE and SED expressions is
difficult. To address this complication, we incorporate a time
dependence in the SE and SED relations, so that we can evaluate these
relations at the time scale of the DH. This point has been neglected
so far in the literature. To define time-dependent versions of the SE
and SED ratios, we first define time-dependent diffusivities
\be
D_t(\Delta t)\equiv\frac{\langle r^2(\Delta t)\rangle}{6\Delta t}\,,\,\,
D_r(\Delta t)\equiv\frac{\langle\varphi^2(\Delta t)\rangle}{4\Delta t}, 
\ee 
and we also define time-dependent relaxation times
\be
\tau(\Delta t)\equiv\int_t^{t+\Delta t}F(q,t')dt'\,. 
\ee
Note that $D_t(\Delta t)\to D_t$ and $D_r(\Delta t)\to D_r$ in the
limit $\Delta t\to\infty$. The definition of $\tau(\Delta t)$ requires
some explanation: $\tau(\Delta t)$ is the time integral of the
intermediate scattering function, and $\tau(\Delta t)$ will be
proportional to the standard relaxation time $\tau$
[Eq.~(\ref{eq:isf})] in the limit $\Delta t\to\infty$.  There is a
constant of proportionality resulting from the stretched exponential
form~\cite{note-1}.  When, instead, a DH is considered, $F_{\rm
subset}(q,\Delta t)$ [see Eq.~(\ref{eq:fqtsubset})] is used in the
computation of $\tau(\Delta t)$.  We choose these definitions since,
in the limit $\Delta t\to\infty$, they converge or are proportional to
the corresponding time-independent definitions.  We will use these
time-dependent quantities to examine time-dependent generalizations of
$R_{\rm SE}$ [Eq.~(\ref{eq:rse})] and $R_{\rm SED}$
[Eq.~(\ref{eq:rsed})].

\subsection{Breakdown time scale}

\begin{figure}
\begin{center}
\includegraphics[scale=0.5]{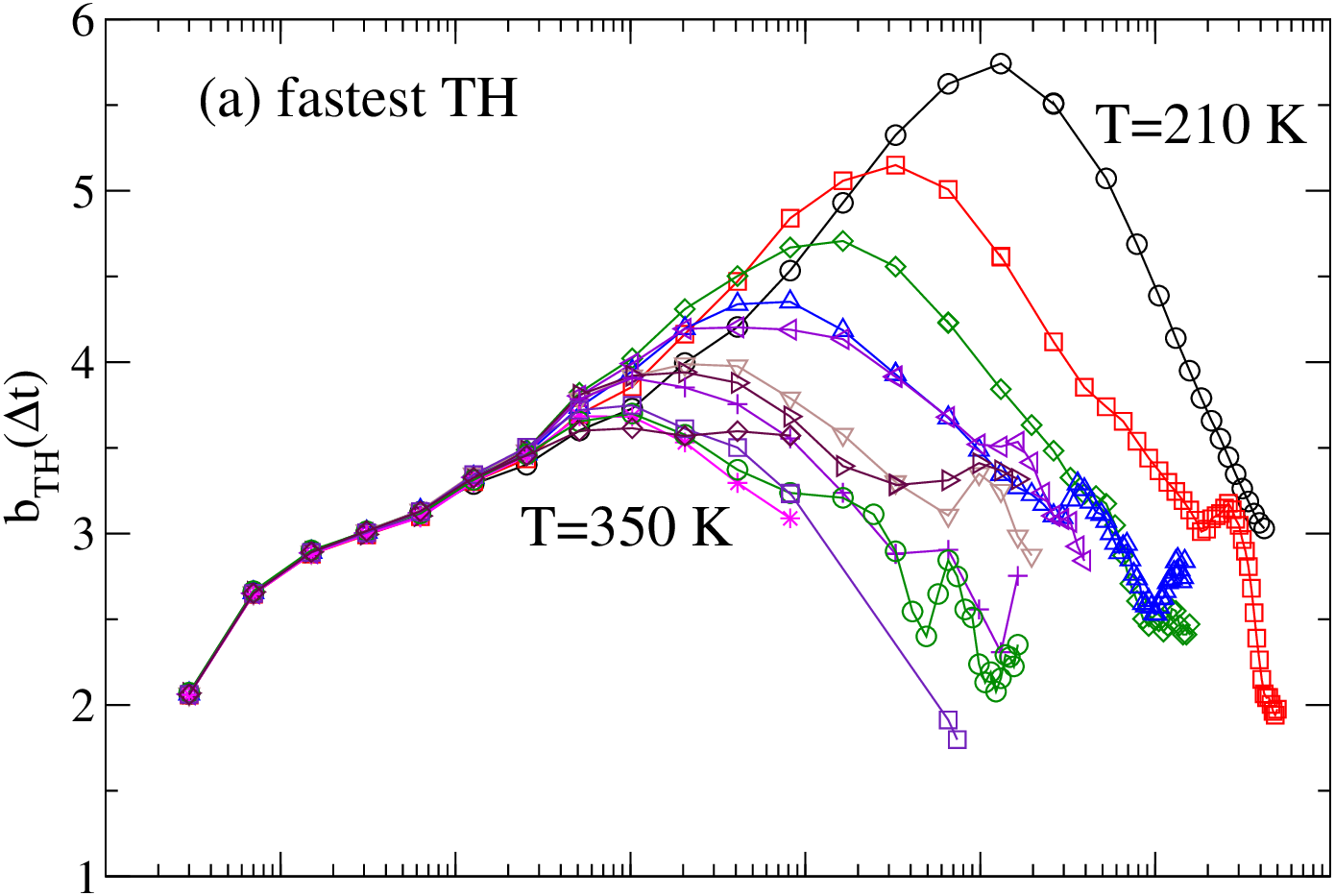}
\includegraphics[scale=0.5]{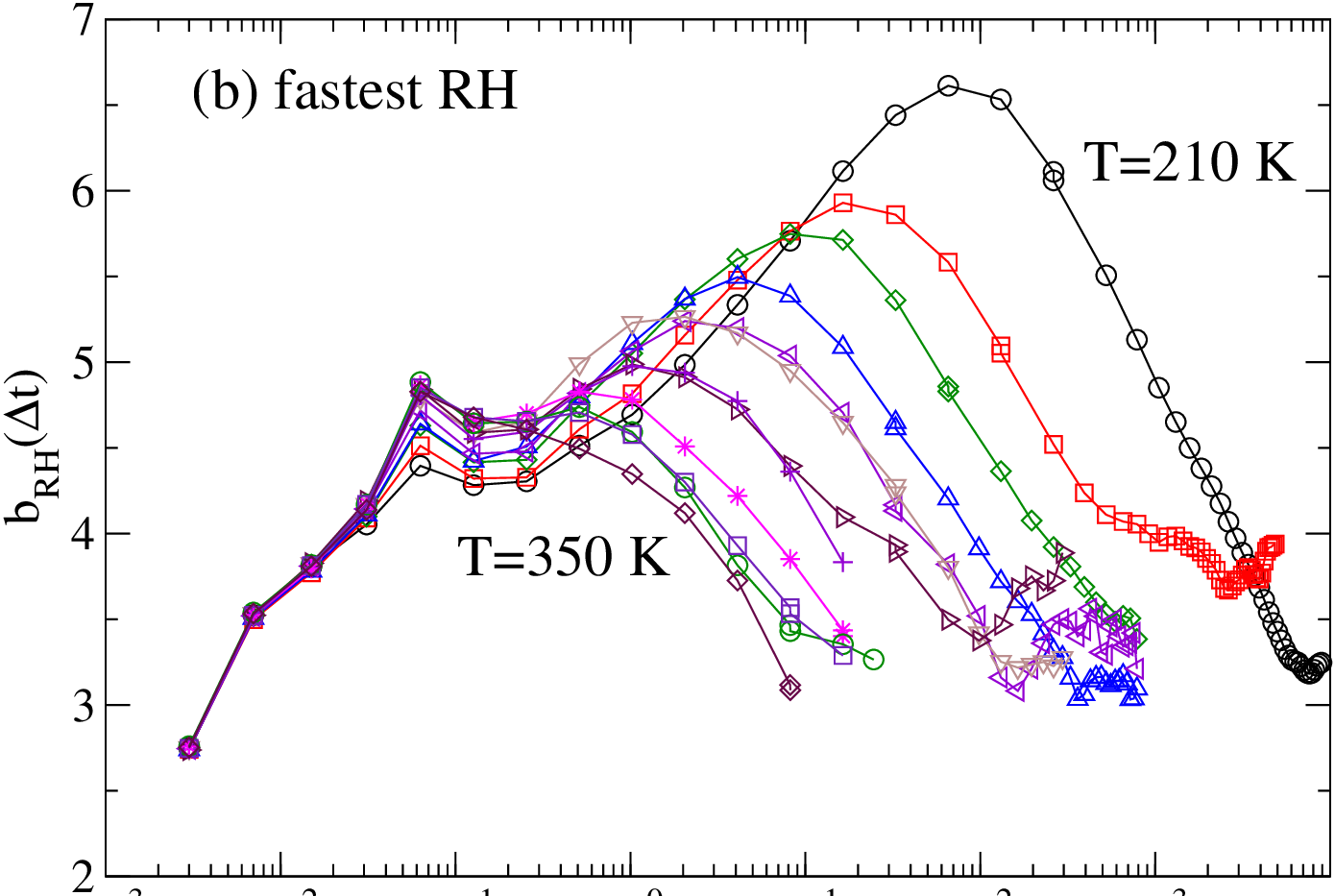}
\end{center}
\caption{(a) Time-dependent extension, $b_{\rm TH}(\Delta t)$, of the
SE relation for the fastest TH at different $T$. For the sake of
clarity the curve corresponding to $T=290$~K was removed. (b)
Time-dependent extension, $b_{\rm RH}(\Delta t)$, of the SED relation
for the fastest RH at different $T$. For the sake of clarity the curve
corresponding to $T=290$~K was removed. In both (a) and (b), the
maxima occur at the time scales corresponding to the end of the cage
regime, when DH are more pronounced. These maxima increase upon
cooling, as the DH become more pronounced.
\label{DtauT-time}}
\end{figure}

Analyzing the time-dependent ratio $D(\Delta t)\tau(\Delta t)/T$ (for
either rotational or translational motion) allows one to verify
quantitatively the role of the time scale in the SE/SED ratios. To
contrast the behavior of $D(\Delta t)\tau(\Delta t)/T$ with the
average over the entire system, we define the time dependent
``breakdown'' ratios as follows:
\be
b_{\rm DH}(\Delta t)\equiv 
\frac{\left(D(\Delta t)\tau(\Delta t)/T\right)_{\rm DH}}
{\left(D(\Delta t)\tau(\Delta t)/T\right)_{\rm all}}
\label{b_DH}
\ee
where DH refers to TH or RH.  If the DH are related to the breakdown
of the SE and SED relations, then one would expect that: (i) the
$b_{\rm TH}$ and $b_{\rm RH}$ ratios will show the largest deviations
from the system average behavior at the time scale when DH are most
pronounced, i.e. approximately at a time which we denote as $t^*$, at
which the non-Gaussian parameter is a maximum (see
Appendix~\ref{sec:times}). (ii) The lower the $T$, the larger the peak
of $b_{\rm DH}$ is (in agreement with the fact that the DH are more
pronounced as $T$ decreases).  Figure~\ref{DtauT-time}(a) for TH and
Fig.~\ref{DtauT-time}(b) for RH, show the behavior of $b_{\rm
DH}(\Delta t)$ for the fastest subset of molecules, for different
temperatures.  Both expectations (i) and (ii) agree with
Fig.~\ref{DtauT-time}.

From Fig.~\ref{DtauT-time} we can extract the time $t_b$ when $b_{\rm
DH}(\Delta t)$ is a maximum. Figure~\ref{tstarDH}(a) shows $t_b$ for
each of the four subsets: TH fastest/slowest and RH
fastest/slowest. If DH play a significant role in the breakdown of the
SE and SED relations, we would expect that the maximum contribution to
the deviation from the SE and SED relations, occurring at $t_b$,
coincides roughly with the ``classical'' measure of the characteristic
time of DH, $t^*$. Comparison of Fig.~\ref{tstarDH}(a) and
Fig.~\ref{tstarDH}(b) for $T<280$~K shows that $t^*$ is slightly larger
than $t_b$ for the slowest DH, while is shorter than $t_b$ for fastest
DH. Nonetheless, $t_b$ and $t^*$ are approximately the same, and so
the largest contribution to the SE/SED ratio is on the time scale when
DH are most pronounced.  This provides direct evidence for the idea
that the appearance of DH is accompanied by the failure of the SE and
SED ratios.

\subsection{Decoupling time scales}

We next directly probe the relation between DH and the decoupling of
$D_r$ and $D_t$. As discussed above, the time scale at which the DH
are observable is much smaller than the time scale at which the system
is considered diffusive.  Therefore, in analogy to the previous
section, we incorporate a time scale in the $D_r/D_t$ ratio so that
we can compare the decoupling between rotation and translation at the
time scale of the DH. To this end we introduce the ratio
\be 
d_{\rm DH}(\Delta t)\equiv {(D_r(\Delta t)/D_t(\Delta t))_{\rm DH}\over
(D_r(\Delta t)/D_t(\Delta t))_{\rm all}},
\label{EqndDH} 
\ee
where DH refers to TH or RH.

Figure~\ref{phi-r}(a) shows the results for $d_{TH}(\Delta t)$ for the
fastest subsets of molecules. For short times, $d_{TH}(\Delta t)$ does
not depend on time and temperature, since in this initial temporal
regime the dynamics at all temperatures is ballistic, i.e., both
$\langle\varphi^2(\Delta t)\rangle$ and $\langle r^2(\Delta t)\rangle$
are approximately linear with $(\Delta t)^2$.  At intermediate times
$d_{TH}(\Delta t)$ develops a distinct maximum which increases in
magnitude and shifts to larger observation times as $T$ is reduced.
The maximum occurs at the time scale where the fastest molecules of
the TH and RH ``break their cages'' and enter the corresponding
diffusive regimes, see Fig.~\ref{tstarDH}(b). Therefore, the results
of Fig.~\ref{phi-r}(a) also suggest that the decoupling between
rotational and translational motion is largest at approximately the
same time scale at which the DH are most pronounced.  We note from
Fig.~\ref{phi-r}(a) that $d_{TH}(\Delta t)<1$, indicating that the
decoupling of rotational and translational motion observed in the
fastest subsets of TH is smaller than that from the average over the
entire system. As we focus in slower subsets of TH for the same $T$,
we observe that the maximum in $d_{TH}(\Delta t)$ decreases at any
given $T$.

Figure~\ref{phi-r}(b) shows $d_{RH}(\Delta t)$ for the fastest subsets
of molecules.  Similar to the behavior of $d_{TH}(\Delta t)$, at short
times $d_{RH}(\Delta t)$ does not depend on time nor temperature;
molecules move ballistically in this regime. The maxima in
$d_{RH}(\Delta t)$ at $\Delta t\approx 0.1$~ps for all temperatures are a
consequence of the librational molecular motion, enhanced in this case
because we are selecting the fastest subset of RH.  At intermediate
times, we observe a broad minimum in $d_{RH}(\Delta t)$ centered at
$\Delta t\approx t^*$; this minimum becomes deeper and shifts to later
times upon cooling, suggesting that the decoupling in the fastest
subset of RH is largest at approximately the same time scale at which
the DH are more pronounced.  The fact that $d_{TH}(\Delta t)$ shows a
maximum at approximately $t^*$, while $d_{RH}(\Delta t)$ shows a
minimum at $t^*$ is because fastest subsets of RH tend to enhance the
rotational motion with respect to the translational motion, while the
opposite situation occurs for the fastest subsets of TH.  We note from
Fig.~\ref{phi-r}(b) that $d_{RH}(\Delta t)>1$, indicating that the
decoupling of rotational and translational motion observed in the
fastest subsets of RH is larger than that found in the average over
the entire system.

In short, the behavior of $d_{TH}(\Delta t)$ and $d_{RH}(\Delta t)$
indicates that the emergence of DH is correlated to the
rotation/translation decoupling, just as it does for the breakdown of
the SE and SED relations.

\section{\label{sec:conc}Summary}

In this work, we tested in the SPC/E model for water (i) the validity
of the SE and SED equations, (ii) the decoupling of rotational and
translational motion, and (iii) the relation of (i) and (ii) to DH. We
found that at low temperatures there is a breakdown of both the SE and
SED relations and that these relations can be replaced by fractional
functional forms. The SE breakdown is observed in every scale of
translational mobility. Similarly, the SED breakdown is observed in
every scale of rotational mobility.  Thus our results support the view
of a generalized breakdown, instead of a view where only the most
mobile molecules are the origin of the breakdown of the SE and SED
relations, embedded in an inactive background where these relations
hold.

We also found that, upon cooling, there is a decoupling of
translational and rotational motion. This decoupling is also observed
in all scales of rotational and translational mobilities.  In
agreement with MD simulations of an OTP model~\cite{OTP}, we find that
an opposite decoupling is observed depending on whether one uses the
rotational diffusivity, $D_r$, or the rotational relaxation time,
$\tau_2$. In the first case, rotational motion is enhanced upon
cooling with respect to the translational motion, while the opposite
situation holds when choosing $\tau_2$. This is particularly relevant
for experiments, where typically only $\tau_2$ is accessible.

We also found that as the decoupling of $D_r/D_t$ increases, the
number of molecules belonging simultaneously to both RH and TH also
increases. This is counter-intuitive since a stronger decoupling would
suggest less overlapping of TH and RH. Therefore we conclude that the
decoupling of $D_r/D_t$ is significant even at the single molecule
level.

We also explored the role of time scales in the breakdown of the SE
and SED relations and decoupling. To do this we introduced time
dependent versions of the SE and SED expressions. Our results suggest
that both the decoupling and SE and SED breakdowns are originated at
the time scale corresponding to the end of the cage regime, when
diffusion starts. This is also the time scale at which the DH are more
relevant.

Our work also demonstrates that selecting DH on the basis of
translational or rotational displacement more strongly biases the
calculation of diffusion constants than the other dynamical
properties. If appropriate care is taken, this should not be
problematic, but it does make apparent that an alternative approach to
identify DH would be valuable. This is especially true when
contrasting behavior of diffusion constants and relaxation times, as
is the case for the SE and SED relations.

\section{Acknowledgments}

We would like to thank S.R. Becker, P.G. Debenedetti, J. Luo,
T.G. Lombardo, P.H. Poole, and S. Sastry for useful discussions. We
thank the NSF for support under grant number CHE-06-16489.

\appendix

\section{\label{sec:rotational-msd}Evaluation of the Rotational Mean
  Square Displacement}

To calculate $D_r$ [Eq.~(\ref{dr})] we consider the behavior of the
normalized polarization vector $\hat{p}_i(t)$ for molecule $i$
(defined as the normalized vector from the center of mass of the water
molecule to the midpoint of the line joining the two hydrogens). The
molecular rotation will cause a rotation of $\hat{p}_i(t)$. A naive
definition of angular displacement as $\hat{p}_i(t) - \hat{p}_i(0)$
would be insensitive to full molecular rotations, since it would
result in a bounded quantity.  Following Ref.~\cite{KKS}, we avoid
this complication by defining the vector rotational displacement in
the time interval $[t,t+\Delta t]$ as
\be
\label{eq:mobiPHI}
\vec{\varphi}_i(\Delta t)\equiv\int_t^{t+\Delta t}\Delta\vec{\varphi}_i(t')dt',
\ee
where $\Delta\vec{\varphi}_i(t')$ is a vector with direction given by
$\hat{p}_i(t')\times\hat{p}_i(t'+ dt')$ and with magnitude given by
$|\Delta\vec{\varphi}_i(t')| \equiv\cos^{-1}\left(\,\hat{p}_i(t')
\cdot\hat{p}_i(t'+dt')\right)$, i.e., the angle spanned by
 $\hat{p}_i$ in the time interval $[t',t'+dt']$.  Thus,
the vector $\vec{\varphi}_i(\Delta t)$ allows us to define a
trajectory in a $3\text{D}$ space representing the rotational motion
of molecule $i$, analogous to the trajectory defined by
$\vec{r}_i(\Delta t)$ for the translational case.  We define,
in analogy to MSD, a {\it rotational} mean square displacement
(RMSD)~\cite{first,lepo2,KKS}
\be
\langle\varphi^2(\Delta t)\rangle \equiv\frac{1}{N}
\sum_{i=0}^N|\vec{\varphi}_i(t+\Delta t)-\vec{\varphi}_i(t)|^2.
\label{rmsd}
\ee
Using this form, we define $D_r$ as given by Eq.~(\ref{dr}), analogous
to the definition of $D_t$.  We have verified that there is no
qualitative difference, in the results of the present work, when the
polarization vector is replaced by the other two principal directions
of the water molecule.

\section{\label{sec:app2} Correlation functions
for Dynamical heterogeneities}

We introduce a MSD, $\langle r^2(\Delta t)\rangle$, for the fastest
and slowest subsets of molecules by limiting the sum in the
Eq.~(\ref{eq:msd}) to the molecules in the corresponding subset. The
different MSDs at $T=210$~K are shown in
Fig.~\ref{fig:msd-and-fqt}(a).  We note that since the most and least
mobile $7\%$ of the molecules will generally vary as a function of
time, the molecules used to calculate $\langle r^2(\Delta t)\rangle$
will change with time; in other words, when a molecule ceases being
part of a DH, it is no longer considered in the computation of the MSD
and the focus is shifted to the new subset of molecules belonging to
the DH considered. Analyzing the $\langle r^2(\Delta t)\rangle$ for
the collection of subsets from most mobile to least mobile has the
advantage that the mean of $\langle r^2(\Delta t)\rangle$ over the
subsets converges to the MSD for the full system. In a similar fashion
the RMSD, $\langle\varphi^2(\Delta t)\rangle$, is calculated also for
the fastest and slowest rotationally mobile molecules
[Fig.~\ref{fig:msd-and-fqt}(b)].

To complement the single particle dynamics determined by $\langle
r^2(\Delta t)\rangle$ and $\langle\varphi^2(\Delta t)\rangle$, we also
evaluate the coherent intermediate scattering function
\be
F(q,\Delta t) \equiv \frac{1}{N\,S(q)} \sum_{j=1}^N e^{-iqr_j(t+\Delta t)}
\sum_{k=1}^N e^{iqr_k(t)}, 
\label{eq:fqt} 
\ee 
where $S(q)$ is the structure factor. $F(q,\Delta t)$ reflects
two-particle temporal correlations instead of single-particle
correlations (as in the case of the MSD). The normalization factors
ensure that $F(q,0)=1$.  In analogy to our analysis of $\langle
r^2(\Delta t)\rangle$, we would like to evaluate the contribution to
$F(q,\Delta t)$ made by subsets of molecules.  Naively, one might
think this can be simply done by limiting the sums in
Eq.~(\ref{eq:fqt}) to solely those molecules within the subset.
However, taking the mean over the subsets of such a quantity will not
recover the complete $F(q,\Delta t)$, since there will be no
information on the cross-correlations between the subsets.  In order
to include these correlations and define a function that, when
averaged over subsets, will return $F(q,\Delta t)$ (as is the case for
MSD and RMSD), we simply limit one of the two sums to the subset,
while the other sum still extends over all molecules. Mathematically,
we define
\be 
F_{\rm subset}(q,\Delta t) \equiv \frac{1}{N_{\rm subset}\,S(q)} \sum_{j=1}^N
e^{-iqr_j(t+\Delta t)} \sum_{k\in subset} e^{iqr_k(t)}.
\label{eq:fqtsubset} 
\ee 
Note that one must make the choice whether to limit the sum to the
subset at time $t$ or $t+\Delta t$; we have found that in practice
there is little, if any, qualitative difference in this choice.  Thus
we measure the correlations between the subset of molecules at time
$t$ with all molecules at time $t+\Delta t$.  Additionally, $F_{\rm
subset}(q,0)$ is not necessarily $1$; forcing this normalization would
not satisfy the desired condition that the mean over subsets returns
the average over all molecules.  In all cases, we evaluate $F_{\rm
subset}(q,\Delta t)$ at $q=18$~nm$^{-1}$, the value of the transferred
momentum at the first maximum of the structure factor where the
relaxation is slowest (except for the $q\rightarrow 0$ limit).
Figure~\ref{fig:msd-and-fqt}(c) and \ref{fig:msd-and-fqt}(d) show
$F(q,\Delta t)$ for all molecules, and for the fastest and the slowest
TH and RH.

At this point, it is important to compare the behavior of $\langle
r^2(\Delta t)\rangle$ and $\langle\varphi^2(\Delta t)\rangle$ with
that of $F(q,\Delta t)$ for the TH and RH subsets.  Since we define
mobility on the basis of displacement, the behavior of $\langle
r^2(\Delta t)\rangle$ and $\langle\varphi^2(\Delta t)\rangle$ for the
subsets are much more strongly affected than $F_{\rm subset}(q,\Delta
t)$ for the subsets. Additionally, $F_{\rm subset}(q,\Delta t)$
includes cross-correlations both within and between subsets that a
single particle definition of mobility does not include.  More
specifically, the data in Fig.~\ref{fig:msd-and-fqt} at $T=210$~K show
that there is roughly two orders of magnitude difference between
$\langle r^2(\Delta t)\rangle$ for the most and least mobile molecules
(and similar difference for $\langle\varphi^2(\Delta t)\rangle$). We
also find that there is roughly also two orders of magnitude
difference between the most and least mobile molecules for $D_t$ and
$D_r$. For higher $T$, the difference is less pronounced.  When we
examine the relaxation of $F(q,\Delta t)$ for the most and least
mobile subsets, we find only a difference of a factor of $\approx$ 2
between the time scales for relaxation.  Therefore --- not
surprisingly --- selecting mobility based on single particle
displacement results in a much stronger effect on diffusion than it
does for collective relaxation phenomena.  This fact is important for
the comparison between this work and a previous work
\cite{bps}.

\section{\label{sec:times}Characteristic time of Dynamical heterogeneities}

\begin{figure}
\begin{center}
\includegraphics[scale=0.5]{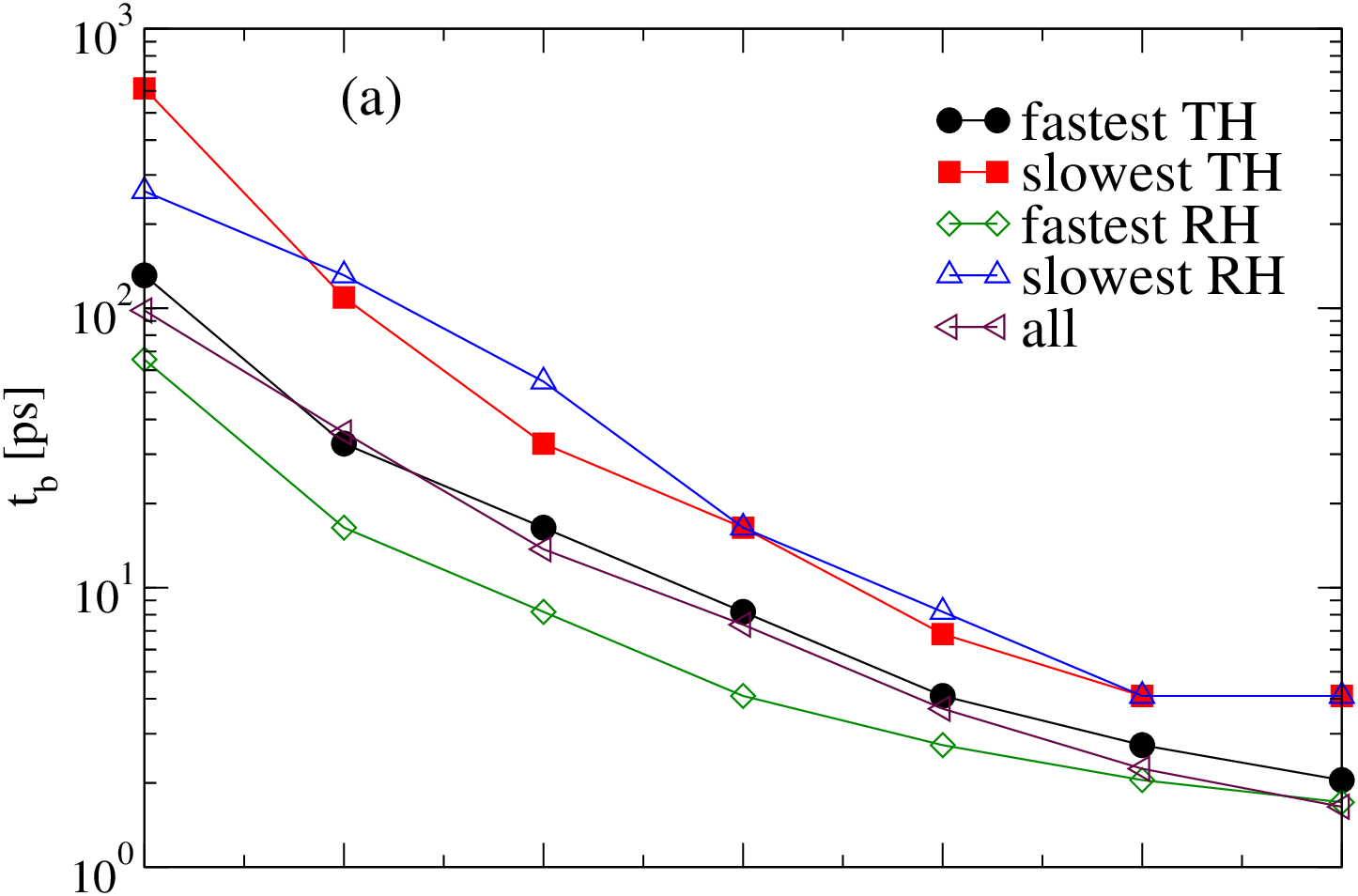}
\includegraphics[scale=0.5]{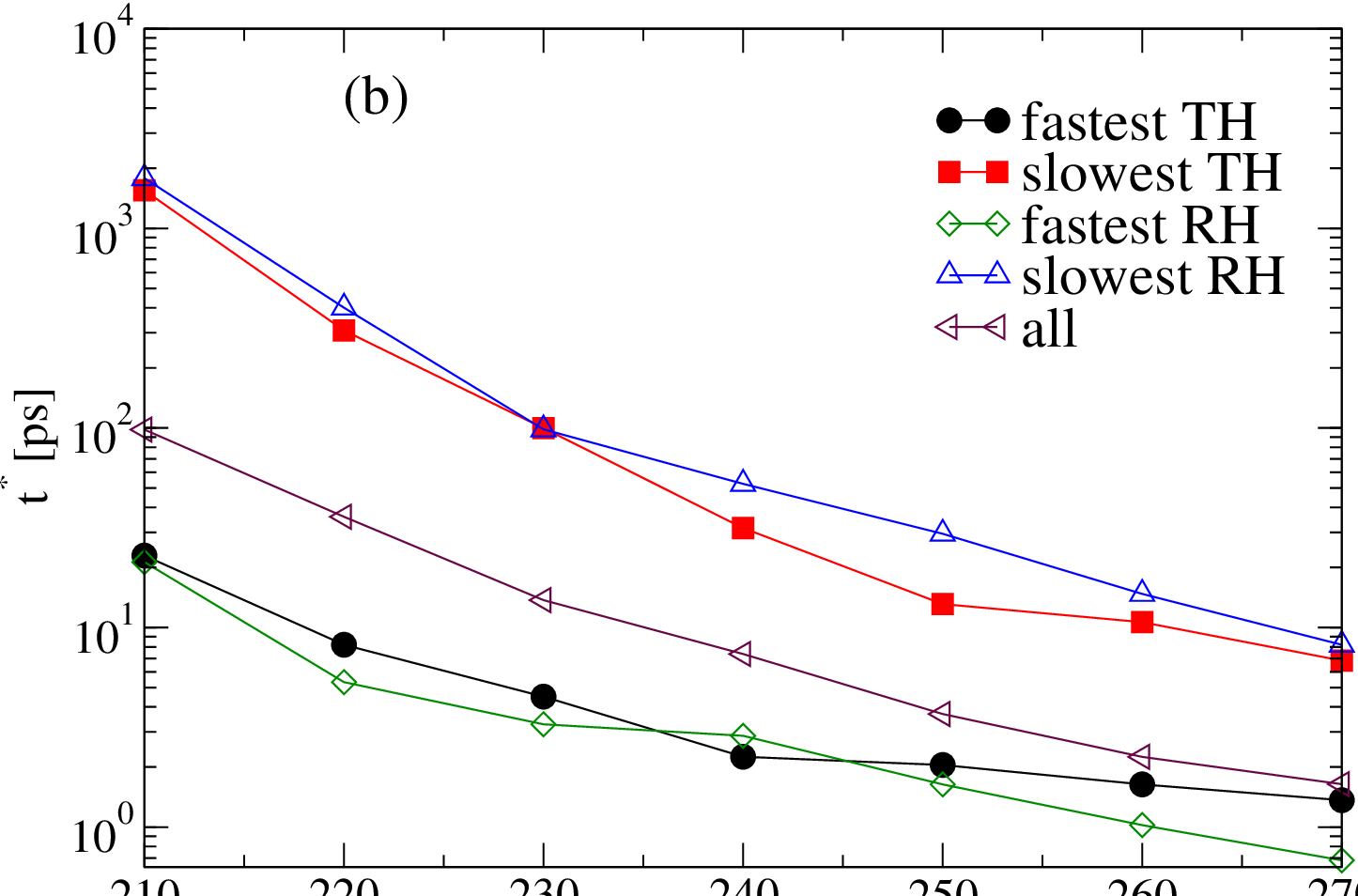}
\end{center}
\caption{Temperature dependence of (a) $t_{b}$, the time at which the
time-dependent extensions of the SE and SED relations, $b_{\rm DH}$,
have maxima, and (b) $t^*$, the time at which the non-Gaussian
parameter, $\alpha_2(\Delta t)$, reaches a maximum. $t^*$ indicates the
time scale corresponding to the end of the cage regime. We show the
results when considering molecules belonging to TH, RH, and also for
the entire system.
\label{tstarDH}}
\end{figure}

\begin{figure}
\begin{center}
\includegraphics[scale=0.5]{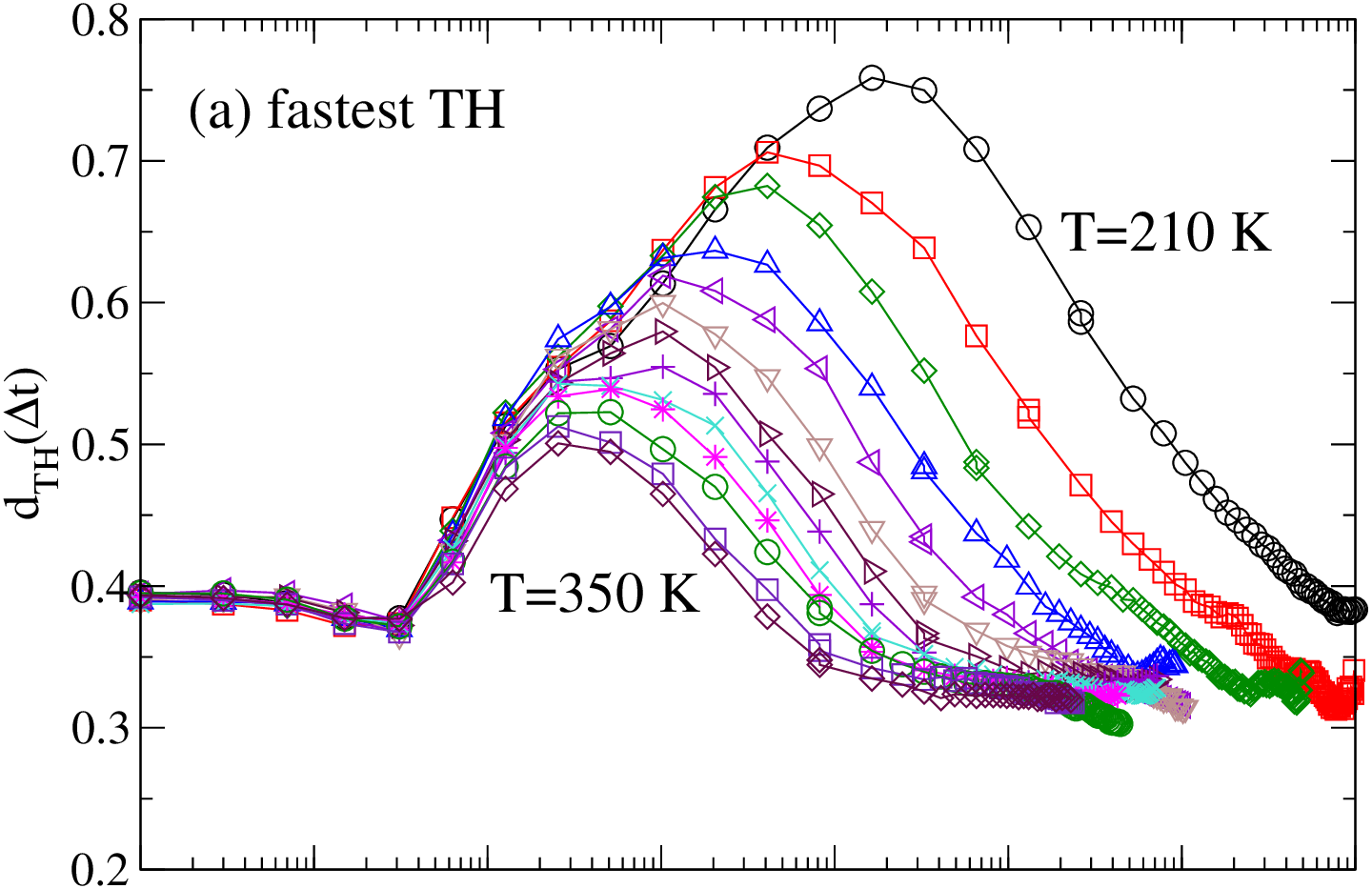}
\includegraphics[scale=0.5]{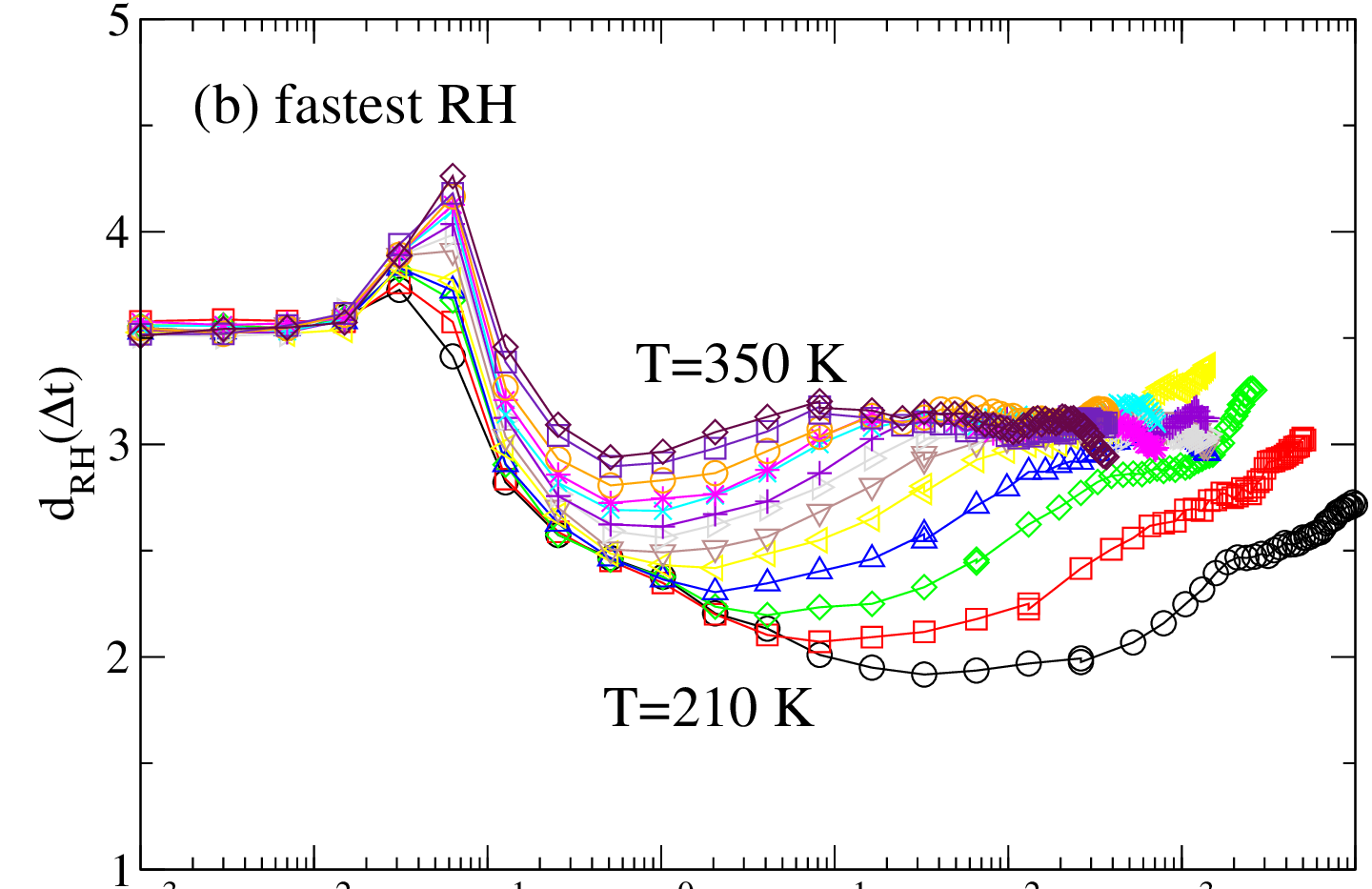}
\end{center}
\caption{(a) Temporal behavior of the ratio of the time-dependent rotational
diffusivity and translational diffusivity for fastest TH, normalized
by the average over the entire system. We show all the simulated
temperatures. (b) Temporal behavior of the ratio of the time-dependent
rotational diffusivity and translational diffusivity for fastest RH,
normalized by the average over the entire system. We show all the
simulated temperatures. The figure shows that the decoupling of
rotation from translation is increasingly more pronounced as $T$
decreases and is a maximum (a) or minimum (b) on the time scale of the
DH.\label{phi-r}}
\end{figure}

\begin{figure}
\centerline{
\includegraphics[scale=0.35]{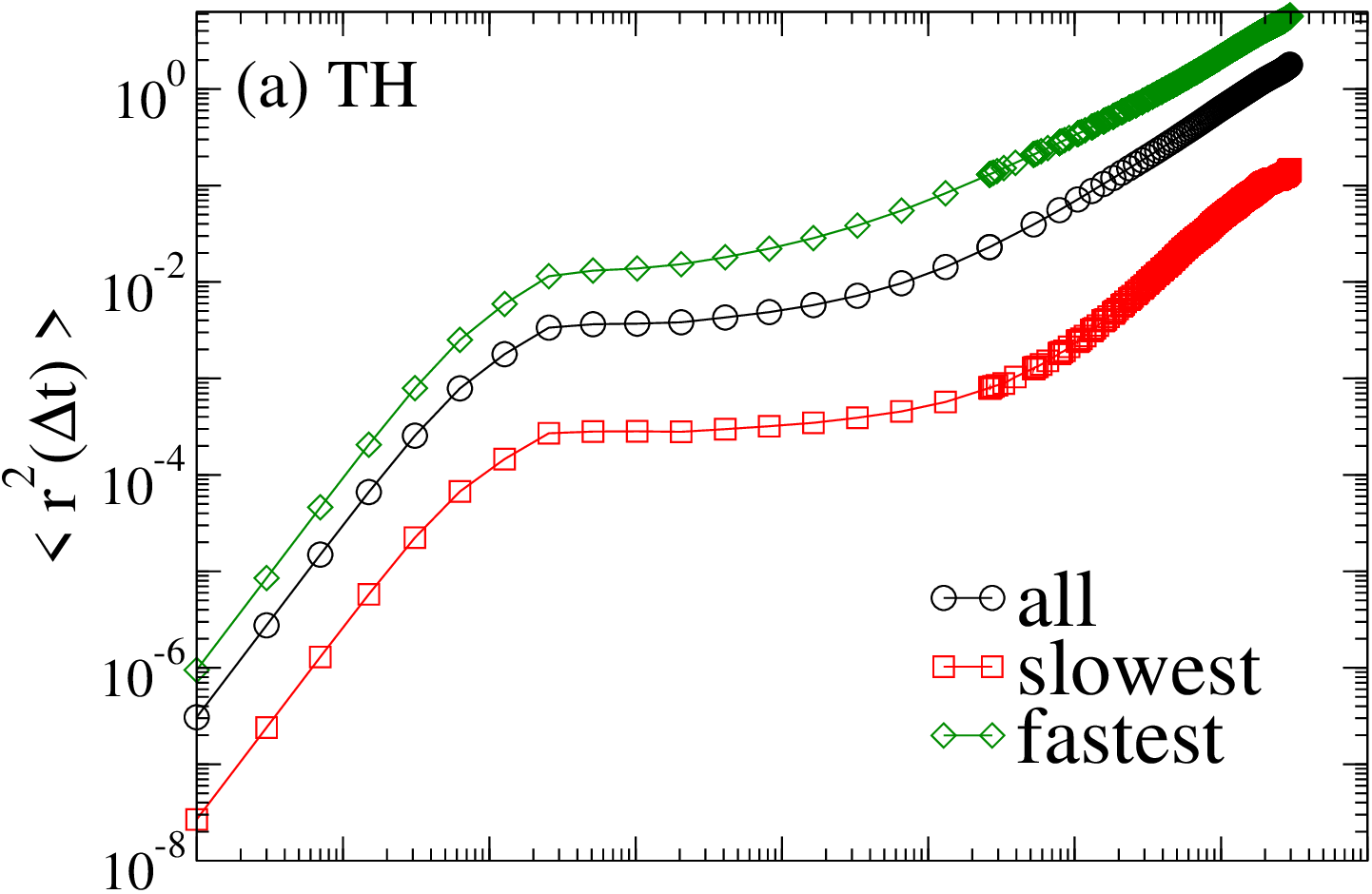}
\includegraphics[scale=0.35]{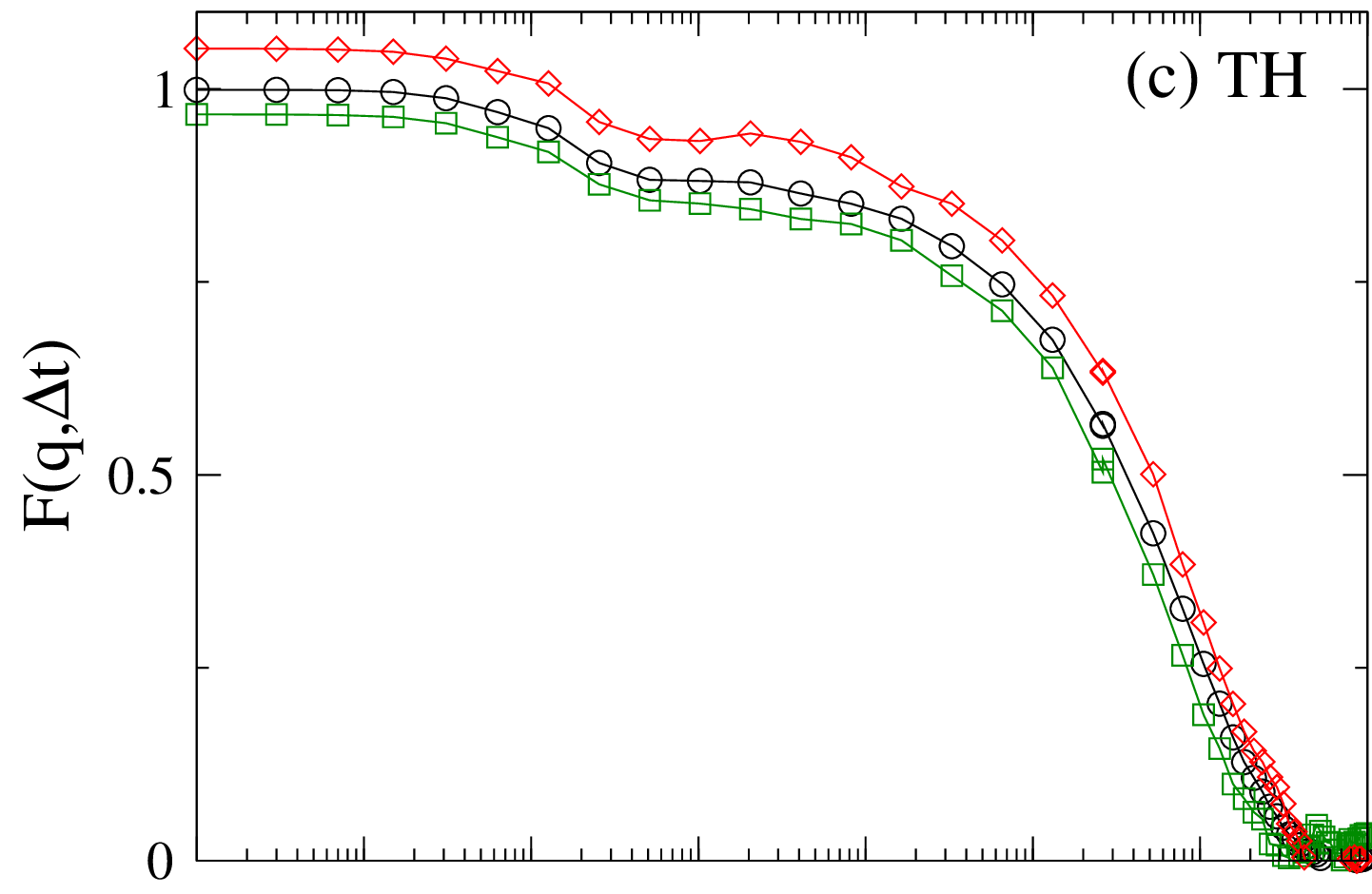}
}
\centerline{
\includegraphics[scale=0.35]{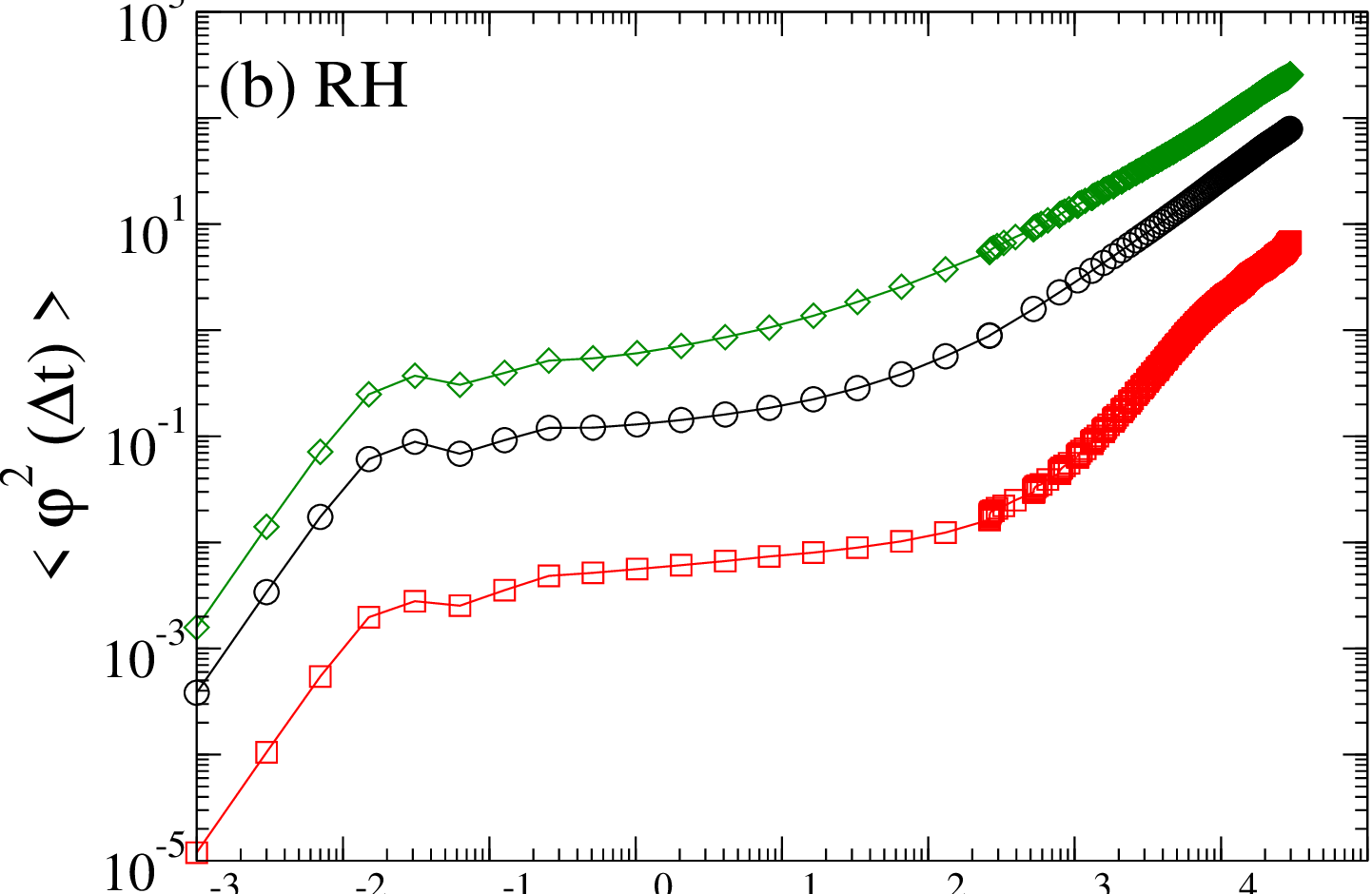}
\includegraphics[scale=0.35]{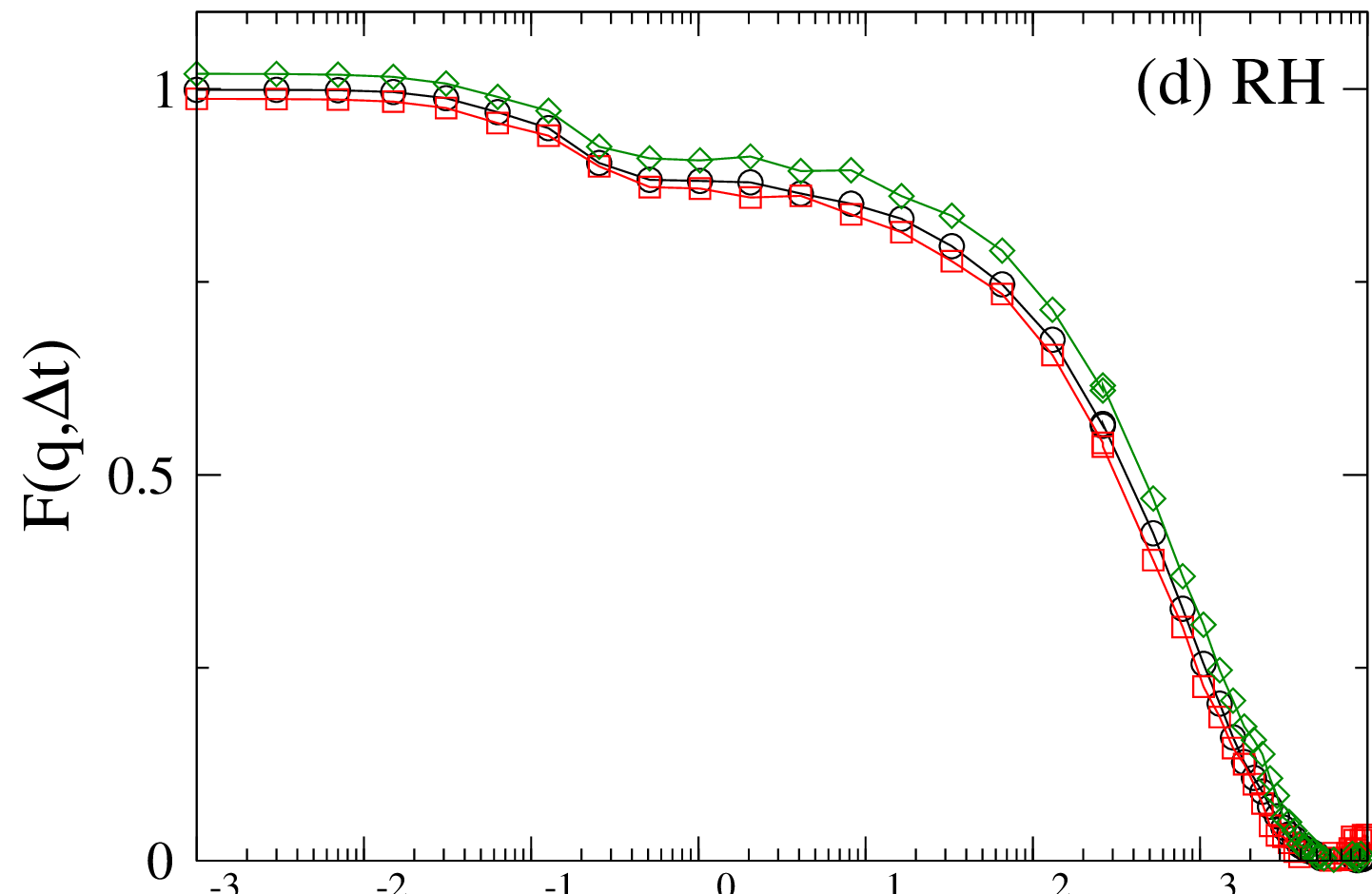}
} 
\caption{Example of time correlation functions limited to subsets of
DH.  (a) MSD and (b) RMSD at $T=210$~K for the fastest and slowest TH
and RH respectively, as well as for the entire system. Intermediate
scattering function, $F(q,\Delta t)$, at $T=210$~K for (c) the fastest
and slowest TH, and entire system and (d) the fastest and slowest RH
and the entire system. \label{fig:msd-and-fqt}}
\end{figure}

Since we analyze the DH both in the context of translational and
rotational motions, it is natural to ask at what time scale the TH and
RH are more pronounced and to what degree the TH and RH subsets
overlap each other. References~\cite{nicoPRL} and~\cite{first} show
that the fastest subsets of TH and RH form clusters, and that these
clusters are larger at approximately the time $t^*$ corresponding to
the onset of the diffusive regime, as indicated by $\langle r^2(\Delta
t)\rangle$ and $\langle \varphi^2(\Delta t)\rangle$
respectively. Normally $t^*$ for the translational case is defined as
the maximum in the non-Gaussian parameter~\cite{rahman}
\be
\alpha_2(\Delta t)\equiv \frac{3<r^4(\Delta t)>}{5<r^2(\Delta t)>}-1\,,
\ee 
where $\langle r^4(\Delta t)\rangle$ and $\langle r^2(\Delta
t)\rangle$ are the fourth and second moment of the displacement
distribution, respectively (the last is also the
MSD). $\alpha_2(\Delta t)$ is known to be identically zero for a
Gaussian distribution, and thus it signals when the dynamics does not
generate such a Gaussian distribution of displacements.  In the
present study, we use either translational, $\vec{r}_i(\Delta t)$, or
rotational, $\vec{\varphi}_i(\Delta t)$, displacement for TH and RH,
respectively, when computing $\alpha_2(\Delta
t)$. Figure~\ref{tstarDH}(b) shows $t^*$ as a function of $T$ defined
for the fastest and slowest subsets of both the TH and RH.  We also
include the corresponding values of $t^*$ for the entire
system. Figure~\ref{tstarDH}(b) shows that there is no qualitative
difference in shape of the curve of $t^*(T)$ for the different subsets
considered and the entire system.

\begin{figure}
\begin{center}
\includegraphics[scale=0.6]{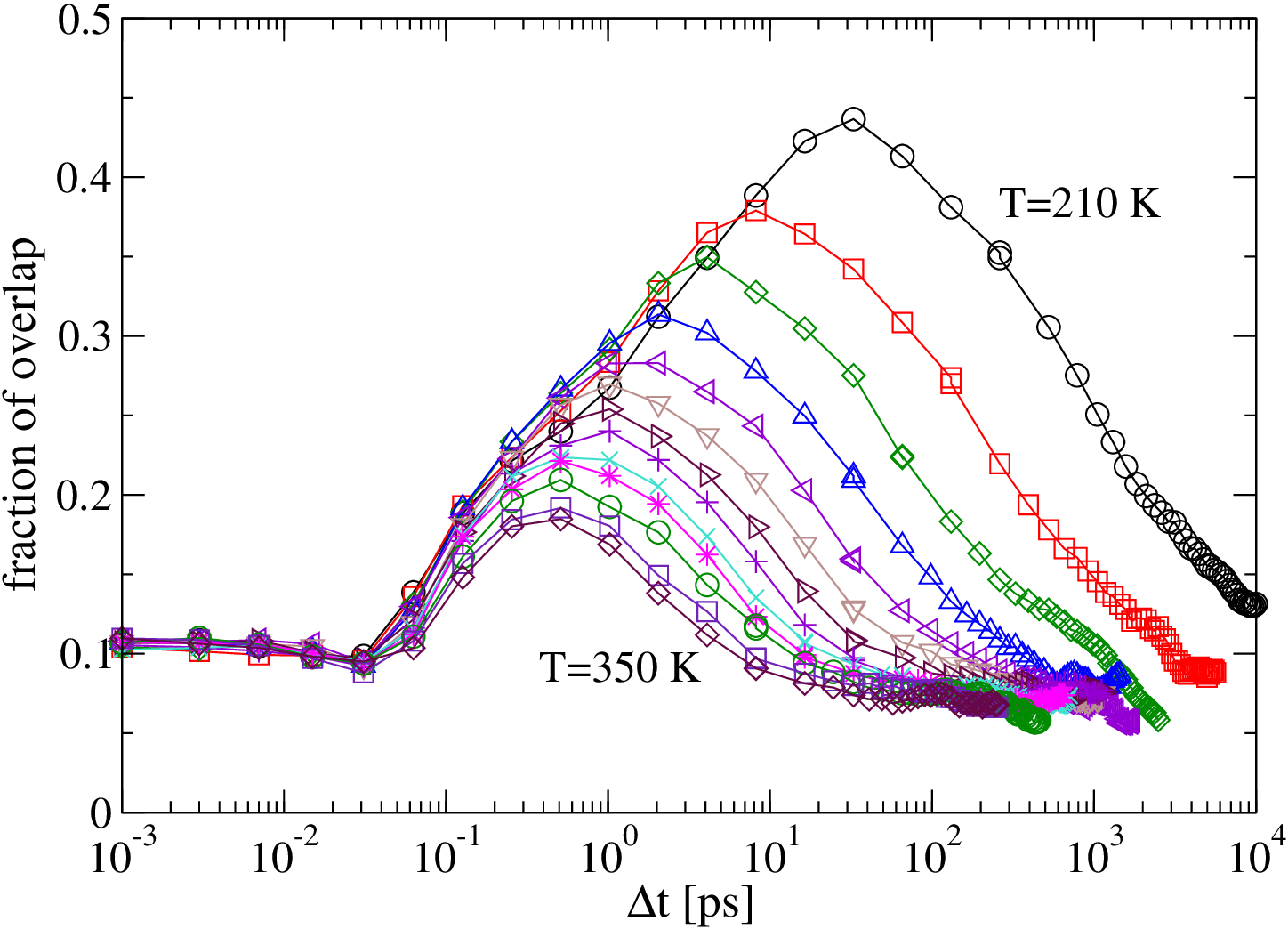}
\end{center}
\caption{Fraction of molecules belonging simultaneously to \emph{both}
fastest TH and fastest RH versus observation time $\Delta t$, at
different temperatures.  This overlap of fastest TH and fastest RH is
maximum at the end of the cage regime and increases upon cooling. It
is almost $45\%$ at the lowest $T$.\label{overlap}}
\end{figure}

Since the values of $t^*$ for TH and RH are similar, we expect
that there is some coupling between TH and RH. Previously, Chen
\emph{et al.} \cite{chengallo} found that there is coupling
between translational and rotational motion at large transferred
momentum $q$. The maximum correlation occurs at the cage relaxation
time, $t^*$, for large values of $q$. Ref.~\cite{first} found a
spatial correlation between RH and TH. Along similar lines, we examine
the overlap between these subsets.  Figure~\ref{overlap} shows the
overlap between the fastest subset of molecules belonging to TH and
RH, as a function of $\Delta t$ and $T$.  Specifically, we count the number
of fastest molecules belonging simultaneously to TH and RH as a
function of observation time $\Delta t$. Similar to Fig. 9 in
Ref.~\cite{chengallo}, the strength of this coupling reaches its
maximum at the cage relaxation times, but these times are consistently
shorter than those reported in \cite{chengallo}; this is likely to be
due to the fact that we consider fastest TH and fastest RH in this
calculation, while Ref.~\cite{chengallo} considers all the molecules
of the system. Figure~\ref{overlap} indicates that, at the lowest
temperature simulated, about $45\%$ of the molecules comprising the
fastest subset of TH coincide with the ones in the fastest subset of
RH.

\eject

\begin{thebibliography}{100}
\bibitem{note-1} Assuming a Kohlrausch form for the intermediate
scattering function, $\exp(-t/\tau)^\beta$, the integral will yield
$\tau(\infty)=\tau \Gamma(\beta^{-1})/\beta$, where $\Gamma(x)$ is the
Euler Gamma function.
\end{thebibliography}
\end{document}